\documentclass[12pt]{article} 
\usepackage{graphicx} 
\usepackage{epsfig}
\usepackage{amsmath, amsfonts, amsthm,amssymb}
\usepackage{latexsym}
\usepackage{xcolor}
\usepackage{mathrsfs}
\usepackage{natbib}
\usepackage{hyperref}
\usepackage{appendix}
\usepackage{dsfont}
\usepackage[left=3cm,top=4cm,right=3cm,bottom=3cm]{geometry}
\RequirePackage[caption=false]{subfig}
\usepackage{lscape}
\usepackage{dsfont}
\usepackage{rotating}
\usepackage{ulem}

\usepackage{mathabx}
\usepackage{calligra}

\newcommand{\e}{\mathds{E}}
\newcommand{\ee}{\mathds{E}}
\newcommand{\dvar}{\mathrm{d}\mathds{V}\mathrm{ar}}
\newcommand{\dcov}{\mathrm{d}\mathds{C}\mathrm{ov}}
\newcommand{\dcor}{\mathrm{d}\mathds{C}\mathrm{or}}

\newcommand{\pdcor}{\mathrm{pd}\mathds{C}\mathrm{or}}
\newcommand{\cov}{\mathds{C}\mathrm{ov}}
\newcommand{\var}{\mathds{V}\mathrm{ar}}
\newcommand{\bY}{\mathbf{Y}}
\newcommand{\bZ}{\mathbf{Z}}
\newcommand{\bC}{\mathbf{C}}

\newcommand{\m}{\mathbf{m}}
\newcommand{\R}{\mathds{R}}

\newcommand{\de}{\mathrm{d}}
\newcommand{\tf}{\mathfrak{t}}

\newcommand{\x}{{\mathbf x}}
\RequirePackage{xcolor}

\graphicspath{{./figs/}}

\begin{document}

\title{\LARGE \bf{Mark distance correlation functions: from moment-based to distributional mark summary characteristics in spatial point processes} }
\maketitle
\begin{center}
{{\bf Matthias Eckardt$^{1}$}, {\bf Mari Myllym{\"a}ki$^{2}$} and {\bf Aila S{\"a}rkk{\"a}$^{3}$}}\\
\noindent $^{\text{1}}$ Chair of Statistics, Humboldt-Universit\"{a}t zu Berlin , Berlin, Germany\\
\noindent $^{\text{2}}$ Natural Resources Institute Finland (Luke) and Department of Mathematics and Statistics,
University of Helsink, Helsinki, Finland\\
\noindent $^{\text{3}}$ Department at Mathematical Sciences, 
Chalmers University of Technology and University of Gothenburg, Gothenburg, Sweden
\end{center} 
\begin{abstract}
With the rapid advancement in data collection devices and storage capacities, we have access to increasing amount of spatial point pattern data where each event location is augmented by multiple, potentially non-scalar marks. Therefore, there is need for efficient analysis techniques to investigate the structural relationships between the marks. In this paper, we recall the distance covariance and distance correlation and adjust them to the marked point process setting. As a result, we introduce a novel class of mark characteristics for single real-valued marks as well as multivariate combinations of marks, including mixtures of integer- and real-valued quantities, and non-scalar marks.
\end{abstract}

\section{Introduction}\label{sec1}
 
Spatial point process data has undergone dramatic changes in size, complexity and dimensionality within recent years. In particular, 
intricate point pattern data, where each point is accompanied by multiple marks of different types, and where the dependence structures between the marks are complex, are becoming more common. Several summary characteristics are available for planar or network-based point patterns with integer or real-valued marks \cite{Illian2008} and have been extended to object-valued, i.e., non-scalar, function-valued marks \cite{Eckardt2023MultiFunctionMarks, Ghorbani2020, Comas2013, Comas2011, comas2008METMA}, composition-valued \cite{EckardtMariCMSPP}, and graph-valued marks \cite{Eckardt:Franaz:Aila}. However, summary statistics for point patterns with multiple diverse marks are still limited. The classical mark correlation function is applied in the situation, where two quantitative marks are attached to a point pattern \cite{Stoyan1987, Raven2011}. Furthermore, patterns with one qualitative and one quantitative mark and bivariate patterns with one quantitative mark have been considered \cite{WiegandMoloney2013}. Apart from using partial spectral coherence functions to unveil conditional dependence structures among multivariate marked spatial point processes through a graphical model approach 
\cite{EckardtISR, Eckardt2018partial, EckardtMateuGonzales2020}, most studies have focused only on pairwise comparisons in the multi-type case. Also, much of the classical literature is limited to linear inter-dependencies, which could lead to skewed results by masking true non-linear dependence.

The classical mark correlation function is fundamentally moment-based. It compares conditional and marginal mark expectations and therefore captures only linear or additive effects in the first moment of the mark distribution. Higher-order dependence, non-monotonic relationships, and changes in the shape of the mark distribution remain invisible to this approach. As a consequence, the moment-based framework can produce misleadingly low correlation values even when strong non-linear associations are present between marks.

To tackle non-linear correlations and complex dependencies, distance covariance and distance correlation which have become valuable tools for assessing dependence between random vectors were introduced in \cite{10.1214/009053607000000505}, sparking a range of methodological innovations and applications across diverse disciplines. Fundamentally, sample distance covariance and distance correlation involve the (adjusted) average product of two double-centered distance matrices. While a zero Pearson correlation coefficient does not necessarily indicate independence and requires specific numerical constraints, zero distance correlation confirms the absolute independence of the involved variables \cite{szelely2023Book}. This framework is particularly robust as it is invariant to shifts in the mark space, requires no positivity assumptions, and is sensitive to any form of stochastic dependence.

Distance correlation has been expanded to include partial \cite{szekely2014}, conditional \cite{cdcor}, and semi-distance correlations \cite{semi}, which accommodate conditional dependence for multivariate random variables with arbitrary dimensions and dependence between a categorical random variable and an arbitrary dimensional random vector as arguments, respectively. Distance correlation has also been applied to mutual independence tests \cite{10.1111/rssb.12259} and feature selection \cite{Tan25012022,  Li01092012} in high-dimensional problems. Although the research has been extended to general metric spaces \cite{10.1214/12-AOP803}, as well as to survival \cite{Zhang+2023, https://doi.org/10.1111/biom.13470, https://doi.org/10.1002/bimj.202200089} and time series analyses \cite{https://doi.org/10.1111/biom.13470, 10.3150:17:BEJ955,  https://doi.org/10.1111/insr.12294}, there is still a notable lack of adaptation to spatial data. Apart from the work by \cite{https://doi.org/10.1111/pirs.12451}, who suggested using a distance correlation index for spatial areal data similar to \cite{https://doi.org/10.1111/j.1538-4632.1985.tb00849.x} to study spatial autocorrelation in election results across various administrative regions, contributions in this area are rare. Specifically, to the best of our knowledge, distance correlation has not yet been applied to (marked) spatial point processes, where the interaction between points and their associated, potentially heterogeneous, attributes must be considered across various spatial scales.

In this paper, we adapt the fundamental ideas of distance correlation \cite{10.1214/009053607000000505} to the marked point process setting and propose a novel mark summary characteristics that enable exploration of variation and possibly non-linear association among marks. We introduce a spatially indexed dependence functional on the conditional pair mark distribution, where distance correlation replaces moment-based summaries as the underlying measure of association. This shift from moment-based to distributional mark characteristics allows us to detect any form of stochastic dependence between marks at a given spatial scale, including non-linear, non-monotonic, and multivariate dependence structures that are invisible to classical mark correlation. The proposed methodology offers a comprehensive framework for statistical description of multidimensional marks across various complexity levels, including (nonlinear) associations for mixed integer- and real-valued data, as well as hybrid mixtures of multivariate object-valued marks. By utilizing a metric-based fusion of distance matrices, our approach allows for the joint assessment of these heterogeneous marks without the need for lossy dimensionality reduction or artificial positivity transformations. Examples of such data can be found e.g.\ in galaxy catalogues, where mass, luminosity, and complex shape descriptors coexist, and extensive forest studies, where trees are characterized by multiple characteristics such as height (real), species (categorical), and crown shape (functional). The proposed framework could in principle also include such shape descriptors as object-valued marks (Section~\ref{sec:dmarkcharacteristics}), although the simulation study and data applications presented here are restricted to quantitative, real-valued, and mixed quantitative-categorical marks. The ability to analyze these hybrid attributes within a single, coherent estimation framework—sensitive to both linear and non-linear interactions—marks a significant advancement in spatial statistics.

The paper is organized as follows. We start by defining marked point processes recalling some of the summary characteristics for them in Section \ref{Preliminaries}. Then, in Section \ref{DistCorr}, we adapt the distance correlation to marked point processes. The proposed methodology is illustrated with simulated data in Section \ref{Simulation} and with real data in Section \ref{Application}. We finish with some concluding remarks in Section \ref{Conclusion}.

\section{Preliminaries}
\label{Preliminaries}

\subsection{Marked point processes}
\label{sec:mpp}

In what follows, let $X = \{x_i, \mathbf{m}(x_i)\}$ denote a simple marked spatial point process with points $x_i \in \mathbb{R}^2$ and multidimensional marks $\mathbf{m}(x_i) = (m_1(x_i), m_2(x_i), \ldots, m_d(x_i))^\top$ living on the space $\mathbb{M}^d = \bigotimes^d_{j=1} \mathbb{M}_j$, where each $\mathbb{M}_j$ is a complete separable metric space equipped with $\sigma$-algebra $\mathcal{M}_j$ and $\mathbb{M}^d$ is equipped with $\sigma$-algebra $\mathcal{M}^d$. For shortness, we also denote $\mathbf{m} = (m_1, \ldots, m_d)$. Simplicity implies that multiple coincident (unmarked) points do not occur. We write $\mathcal{X} = \{x_i\}$ to refer to the unmarked ground process associated with $X$, and denote the Borel $\sigma$-algebra in $\mathbb{R}^2$ by $\mathcal{B}$. A realized point pattern of $X$ is denoted by ${\x} = \{x_i, \mathbf{m}(x_i)\}_{i=1, \ldots, n}$.  

The process $X$ is said to be stationary if $\{x_i, \mathbf{m}(x_i)\} \stackrel{d}{=} \{x_i + s, \mathbf{m}(x_i)\}$ for all $s \in \mathbb{R}^2$, meaning that only the locations are transformed and the marks are left unchanged. 
Furthermore, $X$ is isotropic if $\{x_i, \mathbf{m}(x_i)\} \stackrel{d}{=} \{\mathfrak{r}x_i, \mathbf{m}(x_i)\}$ 
for any rotation $\mathfrak{r}$ around the origin. For a stationary process $X$, the expected number of points in a set $B \in \mathcal{B}$ with marks in $L \in \mathcal{M}^d$ is defined as
\[
\Lambda(B \times L) = \lambda \nu(B) M^d(L),
\]
where $\lambda$ is the intensity of the ground process, $\nu(\cdot)$ is the Lebesgue measure, and $M^d$ is a probability measure on $[\mathbb{M}^d, \mathcal{M}^d]$. If each component $m_j$ in $\mathbf{m}$ is a real-valued quantity, $M^d$ is completely determined by the joint mark distribution function $F_{M^d}(\mathbf{m}) = F_{M^d}(m_1, \ldots, m_d) = M^d((-\infty, m_1], \ldots, (-\infty, m_d])$ for $-\infty < m_j < \infty$, $j=1,\ldots,d$. This distribution is given by
\[
F_{M^d}(m_1, \ldots, m_d) = \int_{-\infty}^{m_d} \cdots \int_{-\infty}^{m_1} f_{M^d}(m_1, \ldots, m_d) \, \mathrm{d}m_1 \cdots \mathrm{d}m_d,
\]
with $f_{M^d}$ denoting the joint mark density function, if it exists. The marginal distribution of a single mark component $m_j$ is then defined as $F_{M_j}(m_j) = M_j((-\infty, m_j]) = \int_{-\infty}^{m_j} f_{M_j}(m) \, \mathrm{d}m$, where $f_{M_j}$ is the respective marginal mark density. Furthermore, the mean $\mu_j$ and variance $\sigma_j^2$ of the $j$-th single mark component $m_j$ are given by
\begin{equation}
    \mu_j = \int_{-\infty}^{\infty} m_j f_{M_j}(m_j) \, \mathrm{d}m_j
\end{equation}
and
\begin{equation}
    \sigma^2_j = \int_{-\infty}^{\infty} (m_j - \mu_j)^2 f_{M_j}(m_j) \, \mathrm{d}m_j,
\end{equation}
respectively. 

Beyond the mark characteristics, the second-order structure of the ground process is described by the second-order factorial moment measure $\alpha^{(2)}(\cdot)$. For sets $B_1, B_2 \in \mathcal{B}$, it is defined as
\begin{eqnarray}\label{eq:factorial:measure}
    \alpha^{(2)}(B_1 \times B_2) &= \mathbb{E} \left[ \sum_{x_1, x_2 \in \mathcal{X}}^{\neq} 
    \mathds{1}_{B_1}(x_1) \mathds{1}_{B_2}(x_2) \right],
\end{eqnarray}
where the symbol $\sum^{\neq}$ denotes the sum over distinct pairs of points and $\mathds{1}_B(\cdot)$ is an indicator function. The density of $\alpha^{(2)}$ with respect to Lebesgue measure is called the second-order product density $\varrho^{(2)}$. Heuristically, for any two points $x_1, x_2 \in \mathcal{X}$, $\varrho^{(2)}(x_1, x_2) \, \mathrm{d}x_1 \mathrm{d}x_2$ can be interpreted as the probability of observing exactly one point of $X$ in each of the infinitesimal areas $\mathrm{d}x_1$ and $\mathrm{d}x_2$. For a rigorous treatment of these foundations, the reader is referred to Illian et al. (2008) and Chiu et al. (2013). In fact, second-order product densities play an important in the estimation of classic mark correlation functions whose unnormalized form is computed by taking the ratio of the product density function of the so-called $\tf_f$-factorial moment measure including a user selected test function $\tf_f$ of the marks, 
\begin{eqnarray}
    \alpha_{\tf_f}^{(2)}(B_1 \times B_2)
&=
\e
\left[
\sum^{\neq}_{\substack{(x_1,m(x_1)),\\ (x_2,m(x_2))~\in X}}
\tf_f(m(x_1),m(x_2))
\mathds{1}_{B_1}(x_1)
\mathds{1}_{B_2}(x_2)
\right],
\end{eqnarray}
 and the second-order factorial moment measure \eqref{eq:factorial:measure}.

\subsection{Summary characteristics for marked point processes}

Here, we briefly recall summary statistics for marked point processes with qualitative, real-valued, and function-valued marks, restricting attention to the material that is either directly relevant for introducing the distance correlation in Section~\ref{DistCorr} or used as a comparator in the simulation study and applications below.

\subsubsection{Summary characteristics for qualitative marks}

When marks represent a countable set of entities, the process is referred to as a multivariate or multitype point process. In this simplest case of marked point process settings, each point is assigned a single integer-valued label $m\in\{1,...,k\}$, categorizing the points into $k$ distinct components. Such patterns are commonly analyzed through cross-type and dot-type extensions of the $K$-function, pair-correlation function, and $J$-function \cite{Ripley1976,Jest1996, van11, MollerWaagepetersenBook, Cronie2016}, or through the mark connection and mark mingling functions, which quantify the conditional probability, respectively the general closeness, of two points at distance $r$ possessing marks $i$ and $j$ \cite{StoyanStoyan1994, Illian2008, HUI2014125, Eckardt:Moradi:currrent, Eckardt2024Rejoinder}.

Since the distance correlation framework of Section~\ref{DistCorr} is built on distances between mark values rather than on a fixed test function, it offers comparatively little over this classical multitype toolkit for a \emph{single} qualitative mark: the distance between two labels reduces to a $0/1$ mismatch, and the resulting distance correlation is closely related in structure to the mark connection function. The situation changes, however, once a qualitative mark is combined with one or more further qualitative or quantitative marks into a mark \emph{vector}. In that case, the Gower distance introduced in Section~\ref{sec:dmarkcharacteristics} accommodates nominal components natively, and the resulting multivariate distance correlation captures joint dependence across mixed categorical-and-continuous mark vectors, exactly as illustrated for species, height, and radial increment in the forestry application of Section~\ref{Application}.

\subsubsection{Summary characteristics for real-valued marks}

Real-valued marks are commonly analyzed through a test function $t_f: \mathbb{M} \times \mathbb{M} \to \mathbb{R}^+$ which quantifies the numerical relationship between the marks of two points located at distance $r$ from each other \cite{PenttinenandStoyan1989, Schlather2001}. For a stationary point process, the $t_f$-correlation function $\kappa_{t_f}(r)$ is defined as
\begin{equation}\label{eq:tfcorr}
\kappa_{t_f}(r) = \frac{\mathbb{E} [ t_f(m(x), m(y)) \mid x, y \in X,\ d(x,y)=r ]}{c_{t_f}},
\end{equation}
where the expectation is conditional on $X$ having points at both $x$ and $y$, and
$c_{t_f}$ is a normalizing factor, typically chosen such that $\kappa_{t_f}(r) = 1$ under the assumption of mark independence. These characteristics are designed to capture specific spatial interactions, such as similarity, as, e.g., the mark variogram
\begin{equation}\label{eq:markvario}
\gamma_{mm}(r) = \frac{1}{2}\, \mathbb{E} \left[ (m(x)-m(y))^2 \mid x, y \in X,\ d(x,y)=r \right],
\end{equation}
or mutual reinforcement (e.g., Stoyan’s product-type mark correlation function $\kappa_{mm}(r)$) \cite{StoyanStoyan1994, Illian2008}.

Point patterns with two qualitative marks $m_1$ and $m_2$ and introduced bi-variate correlation between the two marks are considered in \cite{Stoyan1987}, and later in \cite{Raven2011}. Then, the mark of point $x$ is $m(x)=(m_1(x), m_2(x))$. The bi-variate correlation is defined be setting $t_f(m(x),m(y))=m_1(x) m_2(y)$ in \eqref{eq:tfcorr}. In addition, point patterns with one qualitative and one quantitative mark as well as bi-variate patterns with one quantitative mark are discussed in \cite{WiegandMoloney2013}. In the first case, the question of interest is whether the quantitative mark shows spatial correlation with the qualitatively marked points. Patterns under the null model can be obtained by keeping the qualitative marks fixed and randomly shuffling the quantitative marks  over all points. In the second case, correlation between the marks of the two patterns is of interest and patterns under the null model can be obtained by random labeling of the marks within each pattern.

In addition to these correlation-type functions, mark-weighted cumulative statistics, such as the $K_{t_f}(r)$-function \cite{pettinen1992forest}, account for the simultaneous correlation between point locations and mark values:
\begin{equation}\label{eq:Kmm}
 K_{t_f}(r) = \frac{1}{\lambda c_{t_f}} \mathbb{E}_u \left[ \sum_{x \in X, u \neq x}  t_f(m(u), m(x)) \mathds{1}\{ d(u,x) \leq r \} 
 \right].
\end{equation}

\subsubsection{Summary Characteristics for Function-Valued Marks}

A more complex class of marked point processes arises when each point $x_i$ is augmented by a non-scalar quantity, such as a function-valued mark $g(x_i)$ or inherently structured data. In function-valued scenarios, where the mark $g(x_i)$ is defined on a functional space $\mathds{F}(\mathcal{T})$ over a domain $\mathcal{T}$, summary characteristics can be established through extended test functions $\tf_f: \mathds{F} \times \mathds{F} \to \R^+$.

Pointwise specifications of these characteristics, such as the $\tf_f$-correlation function $\kappa_{\tf_f}(r, t)$, analyze the relationship between functions $g_x(t)$ and $g_y(t)$ at a fixed argument $t \in \mathcal{T}$ for points separated by distance $r$ \cite{Eckardt2023MultiFunctionMarks}:
\begin{equation}\label{eq:tfcorr:ftc}
    \kappa_{\tf_f}(r,t) = \frac{\ee_{x,y} [ \tf_f(g_x(t), g_y(t)) |d(x,y)=r ]}{c_{\tf_f}(t)}.
\end{equation}

Typical examples include the pointwise mark variogram $\gamma_{gg}(r,t)$ and Stoyan's mark correlation function $\kappa_{gg}(r,t)$, which utilize pointwise functional means $\mu_g(t)$ and variances $\sigma_g^2(t)$ as normalizing factors. To capture the overall pairwise interrelation between function-valued marks, global characteristics are constructed by integrating these pointwise versions over the domain $\mathcal{T}$. For instance, the global mark variogram $\gamma_{gg}(r) = \int_{\mathcal{T}} \gamma_{gg}(r, t) \de t$ provides the average variability across pairs of function values in an $\mathcal{L}_2$ sense.

Similarly, global versions of Stoyan's and other classical mark correlation-type functions allow for a comprehensive interpretation of spatial functional dependence, where values typically converge to 1 under mark independence. These integrative concepts also extend to mark-weighted $K$-functions.

Finally, the framework accounts for multivariate settings through cross-type, cross-function, and cross-type cross-function characteristics. These extensions investigate the interrelation of different functions $g_1(\cdot) \neq g_2(\cdot)$ or the dependencies between marks of points belonging to distinct sub-processes $X_i$ and $X_j$. For high-dimensional scenarios, multi-function versions facilitate the analysis of sets of $s$ bundled functions $\mathbf{g}(x_i) \in \mathds{F}^s$ associated with each location \cite{comas2008METMA, Eckardt2023MultiFunctionMarks}.

\subsection{Distance covariance and correlation}
\label{DistCorr}

To quantify dependence between random vectors beyond linear relationships, distance correlation introduced by \cite{10.1214/009053607000000505} can be used. Below, we recall the definition of distance correlation and partial distance correlation. 

\subsubsection{Distance covariance and correlation for random vectors}\label{sec:dcor}

To quantify dependence between random vectors beyond linear relationships, we utilize the concept of distance correlation introduced by \cite{10.1214/009053607000000505}. Let $\bY \in \R^p$ and $\bZ \in \R^q$ be random vectors with finite first moments, where the dimensions $p$ and $q$ do not need to be equal. The distance covariance, denoted as $\dcov(\bY, \bZ)$, was originally motivated by the weighted $L_2$-distance between the joint characteristic function $\varphi_{\bY, \bZ}$ and the product of the marginal characteristic functions $\varphi_{\bY}\varphi_{\bZ}$. While this representation establishes the theoretical link to independence, the distance covariance is most effectively computed in the spatial domain using Euclidean distances. Specifically, the squared distance covariance is defined as
\begin{equation}\label{eq:dcov}
\dcov^2(\bY, \bZ) = \e\left[\Vert \bY - \bY' \Vert \Vert \bZ -\bZ'\Vert \right]
+ \e\left[\Vert \bY-\bY'\Vert \right]\e\left[\Vert \bZ - \bZ''\Vert \right]
- 2\e\left[\Vert \bY-\bY' \Vert \Vert \bZ - \bZ'' \Vert \right],
\end{equation}
where $(\bY', \bZ')$ and $(\bY'', \bZ'')$ are independent copies of $(\bY, \bZ)$. The squared distance variances $\dvar^2(\bY) = \dcov^2(\bY, \bY)$ and $\dvar^2(\bZ) = \dcov^2(\bZ, \bZ)$ are defined analogously, with $\dvar(\bY) = \sqrt{\dvar^2(\bY)}$ referred to as the distance standard deviation of $\bY$. The distance correlation coefficient, $\dcor(\bY, \bZ)$, is then obtained by standardizing the distance covariance by the product of the marginal distance standard deviations, in direct analogy to the classical Pearson correlation $\rho = \cov(Y,Z)/\sqrt{\var(Y)\var(Z)}$. The squared distance correlation is given by
\begin{equation}\label{eq:dcor}
  \dcor^2(\bY, \bZ) =
  \begin{cases}
    \frac{\dcov^2(\bY, \bZ)}{\sqrt{\dvar^2(\bY)\dvar^2(\bZ)}}, & \dvar^2(\bY)\dvar^2(\bZ) > 0 \\
    0, & \dvar^2(\bY)\dvar^2(\bZ) = 0.
  \end{cases}
\end{equation}
The distance correlation is bounded as $0 \leq \dcor(\bY, \bZ) \leq 1$, where $\dcor(\bY, \bZ) = 0$ if and only if $\bY$ and $\bZ$ are independent.

Given $n$ i.i.d.\ copies $(\bY_1,\bZ_1), \ldots, (\bY_n,\bZ_n)$ of $(\bY,\bZ)$, the sample (empirical) distance correlation, denoted $\mathcal{R}_n(\bY, \bZ)$, is obtained by replacing every population expectation in \eqref{eq:dcov}--\eqref{eq:dcor} by the corresponding U-statistic based on the $n\times n$ pairwise distance matrices of the sample \cite{10.1214/009053607000000505}. We use $\mathcal{R}_n(\cdot,\cdot)$ throughout Section~\ref{sec:dmarkcharacteristics} as the empirical distance correlation functional underlying the estimation of the proposed mark distance correlation function.

\subsubsection{Partial distance correlation}\label{sec:pdcor}
While distance correlation provides a powerful tool to detect general (nonlinear) dependence between two random vectors, it does not distinguish between direct and indirect dependencies induced by a third variable. In many applications, however, it is of interest to quantify the remaining association between two variables after accounting for the influence of an additional random vector. This motivates the use of partial distance correlation \cite{szekely2014}.

Let $\bY \in \R^p$, $\bZ \in \R^q$, and $\bC \in \R^r$ be random vectors with finite first moments. The partial distance correlation between $\bY$ and $\bZ$ given $\bC$, denoted by $\pdcor(\bY, \bZ |\bC)$, measures the strength of dependence between $\bY$ and $\bZ$ after removing the effect of $\bC$.

Analogously to classical Pearson partial correlation, one might formally construct the sample partial distance correlation coefficient directly from the pairwise distance covariances. For distance covariances 
$\dcov(\bY,\bZ)$, $\dcov(\bY,\bC)$ and $\dcov(\bZ,\bC)$, and distance variance $\dvar(\bC)$ with $\dvar(\bC) > 0$, the empirical partial distance correlation can be computed by
\begin{equation}\label{eq:pdcor_analogue}
R^*(\bY, \bZ |\bC) =
\frac{\dcov(\bY,\bZ) - \dcov(\bY,\bC)\dcov(\bZ,\bC)/\dvar(\bC)}
{\sqrt{\left(\dvar(\bY) - \dcov^2(\bY,\bC)/\dvar(\bC)\right)
\left(\dvar(\bZ) - \dcov^2(\bZ,\bC)/\dvar(\bC)\right)}}.
\end{equation}
However, as shown by \cite{szekely2014}, this direct algebraic extension generally does not preserve the structural properties required for a true partial correlation of the population. Instead, the rigorous definition of partial distance correlation is based on a Hilbert space formulation involving doubly centered distance functions.

Let $a_{\bY}(\bY,\bY') = \|\bY - \bY'\|$ and define its doubly centered version
\[
A_{\bY}(\bY,\bY') = a_{\bY}(\bY,\bY') - \e[a_{\bY}(\bY,\bY') |\bY] - \e[a_{\bY}(\bY,\bY') |\bY'] + \e[a_{\bY}(\bY,\bY')],
\]
which satisfies $\e[A_{\bY}(\bY,\bY') |\bY] = 0$ and $\e[A_{\bY}(\bY,\bY') |\bY'] = 0$ and $A_{\bZ}$ and $A_{\bC}$ are defined analogously. These doubly centered distance functions are elements of a Hilbert space with inner product
\[
\langle A_{\bY}, A_{\bZ} \rangle = \e\big[ A_{\bY}(\bY,\bY') A_{\bZ}(\bZ,\bZ') \big],
\]
where $(\bY,\bZ)$ and $(\bY',\bZ')$ are independent copies, so that $\dcov^2(\bY,\bZ) = \langle A_{\bY}, A_{\bZ} \rangle$.

The true population partial distance correlation is then defined by projecting $A_{\bY}$ and $A_{\bZ}$ onto the orthogonal complement of the subspace generated by $A_{\bC}$. Denoting the corresponding residuals by
\[
\widetilde{A}_{\bY} = A_{\bY} - \frac{\langle A_{\bY}, A_{\bC} \rangle}{\langle A_{\bC}, A_{\bC} \rangle} A_{\bC}, \quad
\widetilde{A}_{\bZ} = A_{\bZ} - \frac{\langle A_{\bZ}, A_{\bC} \rangle}{\langle A_{\bC}, A_{\bC} \rangle} A_{\bC},
\]
the partial distance correlation $\pdcor(\bY, \bZ |\bC)$ is defined as the cosine of the angle between these residuals,
\begin{equation}\label{eq:pdcor_hilbert}
\pdcor(\bY, \bZ |\bC) =
\frac{\langle \widetilde{A}_{\bY}, \widetilde{A}_{\bZ} \rangle}
{\sqrt{\langle \widetilde{A}_{\bY}, \widetilde{A}_{\bY} \rangle 
\langle \widetilde{A}_{\bZ}, \widetilde{A}_{\bZ} \rangle}}.
\end{equation}
This formulation makes it explicit that partial distance correlation corresponds to the correlation of residuals in a Hilbert space of doubly centered distance functions, in direct analogy to partial correlation in Euclidean space. In particular, $\pdcor(\bY, \bZ |\bC) = 0$ if and only if the residual components $\widetilde{A}_{\bY}$ and $\widetilde{A}_{\bZ}$ are orthogonal.

\section{Mark distance characteristics}\label{sec:dmarkcharacteristics}

Here, we adapt the distance correlation and the partial distance correlation to marked spatial point processes $X = \{x_i, \m(x_i)\}$.

\subsection{Mark distance correlation}\label{sec:mmarkdcor}
 
As in Section \ref{sec:mpp},
we consider mark characteristics defined in terms of
the two-point mark distribution of marks $\m(x)$ and $\m(y)$ 
connected to points $x\in \mathcal{X}$ and $y\in \mathcal{X}$ at distance $r$ apart. We define the mark distance covariance as 
\begin{equation*}
\kappa_{\mathcal{V}}(r) = \dcov^2(\m(x), \m(y)
|x, y \in X,\ d(x,y)=r)
\end{equation*} 
and the mark distance correlation function as 
\begin{equation}\label{eq:dCor}
\kappa_{\mathcal{R}}(r) = \text{dcor}(\m(x), \m(y) |x, y \in X,\ d(x,y)=r).
\end{equation}
For scalar marks ($d=1$), the Euclidean norm $\|\cdot\|$ on $\R^p$ appearing in \eqref{eq:dcov} reduces to the absolute value $|\cdot|$ on $\R$, so that $\kappa_{\mathcal{V}}(r)$ and $\kappa_{\mathcal{R}}(r)$ follow directly from \eqref{eq:dcov} and \eqref{eq:dcor} with $p=q=1$.

The traditional mark correlation function $k_{mm}(r)$ is normalized by the squared mean of the marks, resulting in a reference value of $k_{mm}(r) = 1$ under uncorrelated marks and independent marking. Distance correlation, $\kappa_{\mathcal{R}}(r)$, on the other hand, is standardized by the square roots of distance variances, leading to a reference value of $\kappa_{\mathcal{R}}(r) = 0$. Furthermore, $\kappa_{\mathcal{R}}(r)$ directly quantifies the strength of deviation from independence on a scale from 0 to 1.  

In the univariate setting ($d=1$), $\kappa_{\mathcal{R}}(r)$ offers significant advantages over traditional product-moment based characteristics. 
Stoyan's mark correlation function $k_{mm}(r)$, which is normalized by the squared mean mark $\mu^2$, should only be used for positive marks. In fact, the $k_{mm}(r)$ function loses its interpretability and might lead to confounding effects if applied to centered or residual data. Further it may be sensitive to anomalies in the data.   
In contrast, the distance-based approach is invariant to translation and remains robust regardless of the absolute numerical sign, as it relies on the relative geometry of the mark space. 

The distance correlation $\kappa_{\mathcal{R}}(r)$ is designed to be sensitive to the full geometry of the mark distribution and captures both the location and the spread of that distribution simultaneously. In particular, $\kappa_{\mathcal{R}}(r) = 0$ implies that the marks at distance $r$ are stochastically independent which makes it a necessary and sufficient condition for detecting any form of non-random structure in the mark distribution, including heteroscedasticity, parabolic coupling, and all other forms of dependence that are invisible to moment-based measures.
On the other hand, the $k_{mm}(r)$ function is fundamentally a first-order summary of the two-point mark distribution and is primarily sensitive to linear forms of dependence. In particular, it detects whether nearby points tend to exhibit marks that jointly deviate from their mean in the same direction.
As a consequence, $k_{mm}(r)$ may fail to detect dependence structures that do not manifest through the product moment. As a canonical example, suppose that marks at nearby locations exhibit similar magnitudes but arbitrary sign. In this situation, positive and negative contributions to the product moment may cancel, yielding values close to those expected under independence despite the presence of a structured dependence pattern. More generally, consider a nonlinear relationship of the form $m(y)\approx m(x)^2+\varepsilon$. Even when $m(x)$ and $m(y)$ are strongly coupled, the corresponding product moment can be small or even vanish for symmetric marginal distributions because positive and negative contributions cancel in expectation. Thus, the product-moment characteristic may remain close to its independence reference value although a pronounced nonlinear dependence is present. Similar effects arise for other nonlinear relationships such as $m(y)\approx |m(x)$ or $m(y)\approx \sin^2(m(x))$, where substantial dependence may exist while the first-moment summary captures little or none of it.

In the multivariate setting ($d > 1$), i.e.\ when each point carries a real-valued mark vector $\mathbf{m}(x) \in \R^d$, $\kappa_{\mathcal{R}}(r)$ facilitates the analysis of complex mark vectors without requiring dimension reduction or the loss of structural information. By utilizing Euclidean distances in $\R^d$, it simultaneously captures dependencies within and between different mark components. This is particularly relevant for capturing non-linear cross-interactions and multi-directional spatial dependencies that remain entirely hidden to classical cross-mark correlation functions, which are often restricted to isolated pairwise comparisons. In particular, recalling that the mark distance characteristic derives directly from the correlation between the within-distances of any two mark vectors at a specific distance bin, alternative distances could be applied to account e.g. for mixed-type mark vectors including quantitative and qualitative mark components. A suitable choice for any such mixed mark scenarios include the Gower distance defined 
by
\[
d(i,j) = \frac{\sum_k{ d_{ijk} w_k}}{\sum_k{ w_k}}
\]
where $d_{ijk}$ is the distance between the $i$-th and $j$-th defined by 
\[
d_{ijk}=\frac{\left|x_{ik}-x_{jk}\right|}{\mathrm{range}(x_k)}
\]
if the $k$-th variable is numerical and 
\[
d_{ijk}=
\begin{cases}
    0,\quad \text{if } x_{ik}=x_{jk}\\
    1, \quad \text{else}
\end{cases}
\]
if $k$-th variable is nominal, and $w_k$ is a weight assigned to variable $k$ \cite{Gower}. 

\subsection{Partial mark distance correlation}\label{sec:pdmarkchar}

We now extend the partial distance correlation framework to marked spatial point processes with real-valued target and control marks. Let $X = \{x_i, \m(x_i)\}$ be a marked point process where each point carries a mark vector $\m(x) = (m_1(x), \ldots, m_d(x))^\top \in \R^d$ with $d \geq 2$. We partition the mark vector into a target component $m_1(x)$ and a control component $\mathbf{c}(x) = (m_2(x), \ldots, m_d(x))^\top \in \R^{d-1}$, and consider points $x$ and $y$ at distance $r$   apart. For a given distance $r$, let $(x_1, y_1), \ldots, (x_n, y_n)$ denote the pairs of points with $d(x_i, y_i) = r$. We collect the target mark components into vectors
\[
\mathbf{M}_x = (m_1(x_1), \ldots, m_1(x_n))^\top \in \R^n, \quad
\mathbf{M}_y = (m_1(y_1), \ldots, m_1(y_n))^\top \in \R^n,
\]
and encode the control components at both sides of each pair into the matrix
\[
\mathbf{C} = 
\begin{pmatrix} 
\mathbf{c}(x_1)^\top & \mathbf{c}(y_1)^\top \\ 
\vdots & \vdots \\ 
\mathbf{c}(x_n)^\top & \mathbf{c}(y_n)^\top 
\end{pmatrix} \in \R^{n \times 2(d-1)}.
\]
The partial mark distance correlation function is then defined as
\begin{eqnarray}\label{eq:pmarkdcor}
 \kappa_{\mathcal{R}}^{\mathrm{par}}(r) = 
\pdcor\!\bigl(\mathbf{M}_x,\, \mathbf{M}_y \mid \mathbf{C},\, d(x,y) = r\bigr),   
\end{eqnarray}
following the Hilbert space construction in Section~\ref{sec:pdcor} with $\bY = \mathbf{M}_x$, $\bZ = \mathbf{M}_y$, and $\bC$ as defined above. Analogously to $\kappa_{\mathcal{R}}(r)$, the reference value of  $\kappa_{\mathcal{R}}^{\mathrm{par}}(r)$ under the null hypothesis of no remaining  dependence is zero. In particular, $\kappa_{\mathcal{R}}^{\mathrm{par}}(r) = 0$  implies that the target mark components $m_1(x)$ and $m_1(y)$ are conditionally independent at distance $r$ after accounting for the joint influence of the control marks $\mathbf{c}(x)$ and $\mathbf{c}(y)$ at both endpoints of the pair. This makes $\kappa_{\mathcal{R}}^{\mathrm{par}}(r)$ particularly useful for  distinguishing direct spatial mark dependence from dependence that is merely induced  by shared variation in the remaining mark components. Extending $\kappa_{\mathcal{R}}^{\mathrm{par}}(r)$ to non-Euclidean target or control marks (e.g.\ functional or compositional, Section~\ref{Conclusion}) follows the same Hilbert space construction of Section~\ref{sec:pdcor}, with the Euclidean norm replaced by the doubly centered distance function of the appropriate metric; we do not develop this extension further here.

\subsection{Note on object-valued marks}\label{sec:objectvaluedmarks}

A significant theoretical advantage of the distance-based framework is its independence from the linear structure of Euclidean space: because $\kappa_{\mathcal{V}}(r)$ and $\kappa_{\mathcal{R}}(r)$ from Sections~\ref{sec:mmarkdcor}--\ref{sec:pdmarkchar} operate exclusively on pairwise distances within a metric mark space $(\mathbb{M}, d)$, they apply unchanged on \emph{any} such space; the only requirement for extending them to complex, object-valued marks is to replace the Euclidean norm $\|\cdot\|$ by the metric $d$ appropriate to that mark space, after which $\kappa_{\mathcal{V}}(r)$ and $\kappa_{\mathcal{R}}(r)$ are computed exactly as before. This flexibility is particularly beneficial for objects that naturally cover negative value ranges or exhibit intricate shape variations, as the framework does not rely on product-moment logic but on geometric similarity.

We illustrate this general principle for the case in which each point $x_i$ is associated with a single such object -- as opposed to a bundle of several objects, discussed below. A primary example is a function-valued mark, where the mark represents a continuous function $f_i(t)$ such as a temporal growth profile, a spectral signature, or a daily temperature curve \cite[see][for details]{comas2008METMA, Ghorbani2020, Eckardt2023MultiFunctionMarks}. For such object-valued marks, the distance-based framework is particularly powerful as it avoids the pitfalls of pointwise integration of classical test functions. In functional data analysis, trajectories often cross the zero-axis or represent fluctuations around a mean, so that integrated versions of traditional mark correlation functions would require all functional values to be positive to remain well-defined. The $L^2$ distance
\begin{equation}
    d(f_i, f_j) = \left( \int_{T} |f_i(t) - f_j(t)|^2 dt \right)^{1/2}
\end{equation}
avoids this requirement entirely and plays exactly the role of the Euclidean norm $\|\mathbf{m}(x) - \mathbf{m}(y)\|$ in Section~\ref{sec:mmarkdcor}: only the scalar pairwise distance $d(f_i, f_j)$ enters the distance covariance/correlation double-centering of Section~\ref{sec:dcor}, with $d(f_i,f_j)$ itself computed directly from the discretised vectors of function values of $f_i$ and $f_j$ (Section~\ref{sec:estimation}). Using this $L^2$-based mark distance correlation allows for the detection of complex spatial dependencies, such as synchronized shape variations, phase-locked interactions, or shape-dependent clustering, regardless of the numerical sign or the linearity of the underlying functional process; we return to a fully worked application of this case in Section~\ref{Application}.

Analogous to the extension from scalar ($d=1$) to vector-valued ($d>1$) real marks in Section~\ref{sec:mmarkdcor}, a single object-valued mark extends naturally to a \emph{bundle} of $s$ such objects attached to each point, e.g.\ a bundle of functions $\mathbf{f}(x_i) = (f_{i,1}(t), \ldots, f_{i,s}(t))^\top$, for which a combined distance $D(\mathbf{f}_i, \mathbf{f}_j)$ can be formed as a weighted sum of the component-wise distances, effectively detecting any form of stochastic dependence between the multiple bundled objects.

Beyond functional marks, composition-valued marks $\mathbf{c}$ (e.g., chemical soil compositions) can be treated analogously, as introduced by \cite{EckardtMariCMSPP}. Since compositional data are constrained to sum to a constant and hence operate on the simplex $\mathbb{S}^d$ rather than on $\R^d$, ordinary Euclidean distance is not an appropriate metric: due to this closure constraint, any two parts of a composition are mechanically forced into a spurious negative correlation under standard Euclidean/product-moment treatment, irrespective of any genuine relationship between them \cite[the ``closure problem'';][]{EckardtMariCMSPP}. Replacing the Euclidean distance by the (log-ratio based) Aitchison distance removes this artefact by respecting the multiplicative geometry of the simplex, so that $\kappa_{\mathcal{V}}(r)$ and $\kappa_{\mathcal{R}}(r)$ built on the Aitchison distance correctly capture genuine relative differences between compositions rather than an artefact of the closure constraint. Finally, since the framework depends only on the pairwise distance matrices of the respective mark spaces, dependencies between entirely heterogeneous mark types (e.g.\ a functional and a compositional mark) can, in principle, be studied through a suitably weighted fusion of their respective metrics. We do not develop the compositional and hybrid-fusion cases further in the present paper, restricting the simulation study and applications below to real-valued, mixed (Gower), and single functional ($L^2$) marks; see Section~\ref{Conclusion} for further discussion.

\subsection{Estimation and statistical inference}\label{sec:estimation}

Let $\mathbf{x} = \{x_1, \dots, x_n\}$ be a realization of a stationary and isotropic point process within an observation window $W \subset \mathds{R}^2$, where each point $x_i$ is associated with a multivariate mark vector $\mathbf{m}(x_i) \in \mathds{M} \subseteq \mathds{R}^d$. To estimate the mark distance correlation function $\kappa_{\mathcal{R}}(r)$, we approximate the scale-dependent dependence structure of the underlying two-point mark distribution through a two-stage procedure involving local sample construction and subsequent functional regularization. First, the domain of interpoint distances is partitioned into disjoint bins $B_k = [r_k, r_{k+1}]$. For each bin, we define the set of index pairs
\begin{eqnarray}
\mathcal{P}_k
=
\left\{(i, j) :
\|x_i - x_j\|
\in
B_k,\ i < j\right\},
\end{eqnarray}
which isolates all point pairs whose Euclidean distance corresponds to the spatial scale defined by the bin $B_k$. Structurally, the cardinality of this set, $|\mathcal{P}_k| = \sum_{i \neq j} \mathbf{1}(\|x_i - x_j\| \in B_k)$, is directly proportional to the empirical unweighted second-order product density $\widehat{\varrho^{(2)}}(r)$ of the point process. The scale-specific mark dependence is estimated by constructing two corresponding matrices, $\mathcal{M}_{i,k}$ containing the marks at the origin and $\mathcal{M}_{j,k}$ containing the marks at interpoint distance $r$, induced by the pairs in $\mathcal{P}_k$
\begin{eqnarray}
\mathcal{M}_{i,k} = \left\{\mathbf{m}(x_i) : (i, j) \in \mathcal{P}_k\right\}, \quad \mathcal{M}_{j,k} = \left\{\mathbf{m}(x_j) : (i, j) \in \mathcal{P}_k\right\}.
\end{eqnarray}

The estimator for the raw mark distance correlation at scale $r_k$ is then defined as the empirical distance correlation functional $\mathcal{R}_n$ of Section~\ref{sec:dcor} evaluated over these induced conditional samples
\begin{eqnarray}
\widehat{k}_{\text{raw}}(r_k) = \mathcal{R}_n(\mathcal{M}_{i,k}, \mathcal{M}_{j,k}).
\end{eqnarray}
In the multivariate setting ($d > 1$), $\mathcal{M}_{i,k}$ and $\mathcal{M}_{j,k}$ are treated as matrices of dimension $|\mathcal{P}_k| \times d$, allowing the statistic to capture arbitrary non-linear dependence structures both within and across the components of the mark vectors.

Theoretically, this estimation framework allows for a direct interpretation within the context of mark correlation functions. While classic unnormalized mark correlation functions $\kappa_{\tf_f}(r)$ are defined as the ratio of a mark-weighted product density function $\varrho_{\tf_f}^{(2)}$ and the unweighted product density $\varrho^{(2)}$, the mark distance correlation function $\kappa_{\mathcal{R}}(r)$ can be conceptualized directly as a ratio of a distance correlation weighted product density function $\varrho^{(2)}_{\mathcal{R}}(r)$ to the unweighted density $\varrho^{(2)}(r)$:
\begin{eqnarray}
k_{\mathcal{R}}(r) = \frac{\varrho^{(2)}_{\mathcal{R}}(r)}{\varrho^{(2)}(r)}.
\end{eqnarray}
In this perspective, the numerator $\varrho^{(2)}_{\mathcal{R}}(r)$ weights the infinitesimal pair probability directly with the standardized distance correlation of the conditional joint pair distribution
\begin{eqnarray}
F_{r}(\de m_1, \de m_2) = \mathbb{P}(M_1 \in \de m_1, M_2 \in \de m_2 \mid d(X_1, X_2) = r)
\end{eqnarray}
This ratio formulation clarifies why the spatial geometry of the point process completely cancels out of the estimator, rendering $k_{\mathcal{R}}(r)$ purely sensitive to conditional mark dependencies without being confounded by local variations in mark variance.

Furthermore, this structural analogy extends to the conditional or partial case. When controlling for confounding marks, the functional $\mathcal{R}_n$ is substituted by the empirical partial mark distance correlation function. Under this setup, the continuous target function represents a ratio of a partial distance correlation weighted product density function to the unweighted product density function, effectively filtering out the structural covariance induced by the conditioning set of marks included:
\begin{eqnarray}
k_{\mathcal{R} \mid C}(r) = \pdcor(M_1, M_2 \mid M_{C\setminus\{1,2\}},  d(X_1, X_2) = r).
\end{eqnarray}

To mitigate sampling variance particularly at larger spatial scales with lower pair densities a locally estimated scatterplot smoothing (LOESS) is applied to the sequence of raw estimates $\{\widehat{k}_{\text{raw}}(r_k)\}$. This nonparametric regression yields a continuous function $\widehat{k}_{\mathcal{R}}(r)$, which can be interpreted as a kernel-weighted smoothing of the empirical scale-dependent dependence functional induced by $F_r$, with the bandwidth adaptively determined by the LOESS span.

As the theoretical distribution of $k_{\mathcal{R}}(r)$ under the null hypothesis of spatial independence is analytically intractable, statistical inference is performed via Monte Carlo methods. We test the random labelling hypothesis, which implies conditional independence between the marks, and marks and the spatial locations. Realizations under this null hypothesis are generated by keeping the spatial configuration of the points $\{x_i\}$ fixed while permuting the mark vectors $\{\mathbf{m}_i\}$ as entire units. This procedure preserves both the spatial geometry of the point pattern and the internal marginal distribution of the marks, while completely removing any spatial coupling.
For each permutation $s \in \{1, \dots, n_{\text{sim}}\}$, the smoothed function $\widehat{k}_{\mathcal{R}}^{(s)}(r)$ is recomputed. To control the family-wise error rate across all spatial scales simultaneously, we construct a global one-sided envelope based on the Extreme Rank Length (ERL) measure \cite{mari1, MrkvickaEtal2020, GETpack}. Because distance correlation is non-negative and theoretically equals zero under complete stochastic independence, the null hypothesis implies  $\mathds{E}[\widehat{k}_{\mathcal{R}}(r)]$  is close to zero but positive up to a small simulation bias. We are therefore exclusively interested in detecting whether the observed distance correlation is significantly greater than expected under random labelling.

Applying the global ERL test to the set of curves $\{\widehat{k}_{\mathcal{R}}(r), \widehat{k}_{\mathcal{R}}^{(1)}(r), \dots, \widehat{k}_{\mathcal{R}}^{(n_{\text{sim}})}(r)\}$ yields a critical upper boundary curve $U_{\text{global}}(r)$. A significant spatial interaction is identified when the observed function exceeds this global envelope at least for a scale $r$ such that
\begin{eqnarray}
\widehat{k}_{\mathcal{R}}(r) > U_{\text{global}}(r).
\end{eqnarray}
Such a deviation indicates a departure from conditional independence in the two-point mark distribution at spatial scale $r$. 

\subsection{On the Interpretation of the mark distance characteristics}\label{sec:interpretation}

The three variants of the mark distance correlation introduced above answer different questions, and it is useful to state this explicitly before turning to the simulated examples and envelope tests. The 
function $\kappa_R(r)$ \eqref{eq:dCor} defined for a single scalar mark $m$, compares $m(x)$ and $m(y)$ for two points at distance $r$ and asks whether 
them are drawn independently from the marginal mark distribution or whether some, possibly nonlinear, coupling exists between them. In this sense it is a direct, form-agnostic competitor to both $\kappa_{mm}(r)$ and $\gamma_{mm}(r)$ for a single mark. The multivariate function, defined for a mark vector $\mathbf{m}(x) = (m_1,\dots,m_d)$, replaces the scalar difference by a distance between mark vectors, chosen according to the mark type: Euclidean for real-valued vectors (Section~\ref{Simulation} and the forestry application below), Gower for mixed quantitative and qualitative components (forestry application), and the $L^2$ distance for functional marks (Section~\ref{Application}); Aitchison and hybrid-fusion metrics for compositional or hybrid object-valued marks (Section~\ref{sec:objectvaluedmarks}) follow the same general construction but are not empirically demonstrated in the present paper. Within this multivariate setting, the auto type version compares the same mark component at both points, so that the multivariate machinery reduces, component by component, to the univariate question above. The cross type version instead compares different mark components across the pair, for instance mark component $A$ at $x$ against mark component $B$ at $y$, and can be computed for every pair of distinct components in the mark vector, yielding a full set of cross curves rather than a single one, as illustrated for all three pairwise combinations of radial increment, height and species in our data example illustrated in Figure~5. The corresponding classical counterpart is not the mark correlation function $\kappa_{mm}(r)$ itself, since that operates on a single mark, but rather the bivariate or cross mark correlation functions in the tradition of Stoyan (1987), obtained by setting the test function to $t_f(m_1(x),m_2(y)) = m_1(x)\,m_2(y)$, or its generalizations to real valued cross marks. These constructions remain moment based, require a prespecified test function, and, as before, are restricted to linear or monotone dependence. Finally, the full joint multivariate version, obtained by feeding all mark components at once into a single distance such as the Gower distance, answers the question whether the complete mark vector at $x$ is associated at all with the complete mark vector at $y$ at distance $r$. Therefore, a significant result at this level, as shown for the three tree marks combined in Figure~3, does not indicate which component or pair of components drives the association; establishing that requires descending to the auto and cross type decomposition shown in Figure~5. The partial version, finally, changes the question from whether dependence exists to whether dependence remains once the variation explained by a specified control mark set $C$ has been removed, and thereby separates direct dependence from dependence that is merely induced by a shared confounder or a mediating mark.

Across all three variants, an exceedance of the upper simulation envelope based on random labeling carries the same basic meaning and the same basic limitation. If $\kappa_R(r)$ exceeds the envelope for at least one $r$, the random labeling hypothesis is rejected: the compared quantities, whether single marks, mark vectors, or residualized marks depending on the variant, are not stochastically independent. This statement carries no information on the direction or the functional form of the dependence, since $\kappa_R(r)$ is nonnegative and form agnostic by construction, unlike $\kappa_{mm}(r)$ or $\gamma_{mm}(r)$, whose sign and magnitude are themselves interpretable. The raw magnitude of $\kappa_R(r)$ should also not be read as an effect size, because the estimator carries a small positive bias even under true independence, only its position relative to the envelope, or equivalently its rank in the global test, is informative, not the numerical value itself.

Since $\kappa_R(r)$ alone cannot reveal the type of dependence it has detected, two complementary steps are needed to characterize it. First, $\kappa_R(r)$ should be compared against $\kappa_{mm}(r)$ and $\gamma_{mm}(r)$ at the same distance $r$. If both the distance based and the classical characteristics exceed their respective envelopes, the dependence contains a component that acts on the first moment, that is, a linear or monotone coupling. If only $\kappa_R(r)$ reacts, the dependence exists but is invisible to the first moments, as is typical of heteroscedasticity, symmetric nonlinear coupling, or relationships whose positive and negative contributions cancel in the product moment. Second, the conditional scatterplot of the mark pairs at the flagged distance should be inspected directly, as done in Figure~4: $\kappa_R(r)$ indicates at which distances such a plot is worth constructing, while the scatterplot itself reveals what form the dependence actually takes, whether parabolic, V shaped, or a widening or narrowing scatter indicative of variance coupling.

The distance profile of $\kappa_R(r)$ itself carries additional, though limited, information. An exceedance confined to small $r$ that decays back into the envelope at larger distances points to a short range, local mark interaction, comparable to reading off a range of interaction from a pair correlation function, as seen in the forestry example in Figure~3. An exceedance that persists across the entire range of $r$ investigated should be treated with more caution, since it may reflect a global trend in the first order mean of the mark rather than a genuine distance dependent interaction, as illustrated by the linear field scenario in Section~4.1. Because $\kappa_R(r)$ is only cleanly interpretable as a function of interpoint distance under approximate stationarity and isotropy of the marking, such a pattern should prompt a check for, and if necessary a removal of, a global mark trend before the curve is read as evidence of spatial interaction.

For the partial version, a further distinction applies. If the unconditional $\kappa_R(r)$ exceeds its envelope while the corresponding partial $\kappa_R^{\mathrm{par}}(r)$ does not, the original association is fully explained by the control marks $C$, corresponding to confounding or mediation, as in the spurious correlation scenario of Section~4.3. If both the unconditional and the partial curve exceed their envelopes, a direct or residual dependence remains that cannot be attributed to $C$ alone. The choice of which mark to place in the control set $C$ still requires subject matter judgement, since conditioning on a collider rather than a genuine confounder can, in principle, induce rather than remove spurious association; the collider scenario considered here did not produce a false positive, but this should not be generalized without further study.

\section{Simulation Study}
\label{Simulation}

To evaluate the performance of the proposed mark distance correlation ($k_{\mathcal{R}}(r)$) compared to standard moment-based approaches, namely the mark correlation function ($k_{mm}(r)$) and the mark variogram ($\gamma_{mm}(r)$), we designed a comprehensive simulation study. The design follows the premise of investigating complex dependence structures under the assumption of stationarity of the unmarked point process $\mathcal{X}$, specifically selecting scenarios that pose a challenge for traditional Pearson-based metrics. It should be noted, however, that in some of the examples below, the marked point process $X$ is not stationary. Although $\kappa_{\mathcal{V}}(r)$ and $\kappa_{\mathcal{R}}(r)$ are formally defined for stationary and isotropic marking (Section~\ref{sec:mmarkdcor}), the associated Monte Carlo test always targets the broader and well-defined null hypothesis of random labelling, i.e.\ marks that are independent and identically distributed across locations; the non-stationary scenarios below are therefore included to probe the statistic's sensitivity under this test and do not admit the same population-level interpretation as $\kappa_{\mathcal{R}}(r)$ under stationarity. All simulations were conducted in a unit square window $W = [0, 1]^2$, with spatial locations $x_i=(x_{i1},x_{i2})$ generated by a homogeneous Poisson point process (CSR) with intensities of $\lambda = 200$ for the univariate and $\lambda = 80$ for the multivariate experiments. 
To evaluate the performance of the proposed distance-based and moment-based summary characteristics we considered in total five distinct scenarios, separated into univariate and multivariate simulation regimes, including independence, linear dependence, and three forms of heteroscedastic mark–space interaction. We simulated 200 patterns and computed $n_{\text{sim}} = 499$ Monte Carlo permutations under the random labeling hypothesis. One-sided envelopes for the mark distance correlation are presented since the values of it are always non-negative. No edge correction was used when estimating the summary characteristics  resulting in a slightly better power of the test \cite{Myllymaki2015}.
The performance is assessed using multiple distributional characteristics of the Monte Carlo p-values, including mean and median, interquartile range (IQR), empirical rejection rates, tail probabilities (corresponding to $p < 0.01$), 
and an area-under-the-curve (AUC) measure summarising overall signal separation.

\subsection{Univariate marks: detection of spatial heteroscedasticity}

The first experiment focuses on the sensitivity to spatial variance structures under a constant expected value which represents a form of  heteroscedasticity frequently encountered in ecological and economic data. The general mark model is defined as $m(x_i) = \mu(x_i) + \sigma(x_i) \cdot \epsilon_i + C$, where $\epsilon_i \sim \mathcal{U}[-0.5, 0.5]$ represents uniform noise and $C=20$ is a base constant to ensure positive mark values for the normalization of $k_{mm}(r)$. 

In the first scenario (independent noise), spatial independence is modeled by $\mu(x_i) = 0$ and $\sigma(x_i) = 1$. A linear trend in the mean is induced in Scenario 2 (linear field) by $\mu(x_i) = 15x_{i1}$ with constant variance $\sigma(x_i) = 1$. The primary differentiation occurs in Scenarios 3 to 5, which represent various forms of heteroscedasticity with a constant mean ($\mu(x_i) = 0$). In Scenario 3 (sinusoidal heteroscedasticity), local volatility follows a periodic grid pattern defined by $\sigma(x_i) = 0.1 + 8 \cdot 
|\sin(4x_{i1})\cos(4x_{i2})|$. Since the local product moment remains constant in expectation, $k_{mm}(r)$ is expected to fail in detecting any structure. The fourth scenario (quadratic variance gradient) extends this to a global variance trend with $\sigma(x_i) = 0.1 + 10x_{i1}^2$, while Scenario 5 (local volatility cluster) models an isolated spatial interaction via a Gaussian kernel: $\sigma(x_i) = 0.1 + 12 \cdot \exp(-\|x_i - (0.5,0.5)\|^2 / 0.03)$. These scenarios test the ability of the characteristics to detect spatial dependence that is expressed through higher-order moments while the first-order mean structure remains constant.  

To evaluate the effectiveness of the proposed mark distance correlation, we compare its performance directly against the classical mark correlation function and mark variogram across  the five univariate scenarios. The results are reported in Table \ref{tab:SimTabRes}. Both classical methods work well under the null scenario and under linear correlation. However, the mark correlation function does not recognize the non-linear correlation scenarious while mark variogram does a slightly better job even though the results are not always significant. The distance correlation, however, recognizes the correlation in each case.  
\begin{table}[ht]
\centering
\begin{tabular}{lccccc}
\hline
Metric & Independent & Linear & Var(Grad) & Var(Sin) & Var(Spot) \\ 
\hline
Mean $p$ ($\gamma_{mm}$)  & 0.52 & 0.00 & 0.08 & 0.19 & 0.09  \\ 
Mean $p$ ($\kappa_{mm}$) & 0.50 & 0.00 & 0.52 & 0.49 & 0.49  \\ 
Mean $p$ ($\kappa_{R}$) & 0.50 & 0.00 & 0.00 & 0.01 & 0.00   \\ 
\hline
Median $p$ ($\gamma_{mm}$)  & 0.52 & 0.00 & 0.01 & 0.11 & 0.07 \\ 
Median $p$ ($\kappa_{mm}$) & 0.50 & 0.00 & 0.49 & 0.45 & 0.51   \\ 
Median $p$ ($\kappa_{R}$) & 0.49 & 0.00 & 0.00 & 0.00 & 0.00  \\ 
\hline
IQR($p$) ($\gamma_{mm}$)& 0.46 & 0.00 & 0.10 & 0.23 & 0.11 \\ 
IQR($p$) ($\kappa_{mm}$) & 0.47 & 0.00 & 0.47 & 0.52 & 0.46   \\ 
IQR($p$) ($\kappa_{R}$) & 0.50 & 0.00 & 0.00 & 0.00 & 0.00   \\ 
\hline
Rejection rate ($\gamma_{mm}$)& 0.05 & 1.00 & 0.60 & 0.30 & 0.55  \\ 
Rejection rate ($\kappa_{mm}$) & 0.06 & 1.00 & 0.01 & 0.04 & 0.04 \\ 
Rejection rate ($\kappa_{R}$)& 0.04 & 1.00 & 1.00 & 0.94 & 1.00 \\ 
\hline
Prop($p<0.01$) ($\gamma_{mm}$)& 0.02 & 1.00 & 0.40 & 0.14 & 0.19 \\ 
Prop($p<0.01$) ($\kappa_{mm}$) & 0.01 & 1.00 & 0.00 & 0.01 & 0.01 \\ 
Prop($p<0.01$) ($\kappa_{R}$) & 0.00 & 1.00 & 1.00 & 0.88 & 1.00  \\ 
\hline
AUC ($\gamma_{mm}$)  & 0.48 & 1.00 & 0.92 & 0.81 & 0.91  \\ 
AUC ($\kappa_{mm}$) & 0.50 & 1.00 & 0.48 & 0.51 & 0.51  \\ 
AUC ($\kappa_{R}$) & 0.50 & 1.00 & 1.00 & 0.99 & 1.00   \\ 
\hline
\end{tabular}
\caption{Comparison of the mark variogram $\gamma_{mm}$, Stoyan's mark correlation $\kappa_{mm}(r)$ and the mark distance correlation functions $\kappa_R(r)$ across all simulation scenarios, each one computed from 200 point patterns. The $p$-values from global envelope tests based on 499 simulations are reported. Heading abbreviations read as Indep = independent noise, Linear = linear field,  Var(Grad) = sinusoidal heteroscedasticity,  Var(Sin) = quadratic variance gradient,  Var(Spot) = local volatility cluster.}
\label{tab:SimTabRes}
\end{table}
Overall, the results suggest that distance-based mark correlation provides a substantially more sensitive and robust framework for detecting complex mark–space dependencies than classical second-order mark correlation functions.  

\begin{sidewaysfigure}
    \centering
        \caption{Results for one simulation run: Marked point patterns (1st row), and their mark distance correlation $\kappa_{\mathcal{R}}(r)$ (2nd row) and classical mark correlation function $\kappa_{mm}(r)$ (3rd row) and mark variogram $\gamma_{mm}(r)$. The envelopes are based on 499 Monte Carlo permutations under random labeling. Headings abbreviation read as Indep = independent noise, Linear = linear field,  Var(Grad) = sinusoidal heteroscedasticity,  Var(Sin) = quadratic variance gradient,  Var(Spot) = local volatility cluster.}
    \label{fig:sim_univariate}
    \includegraphics[scale=.409]{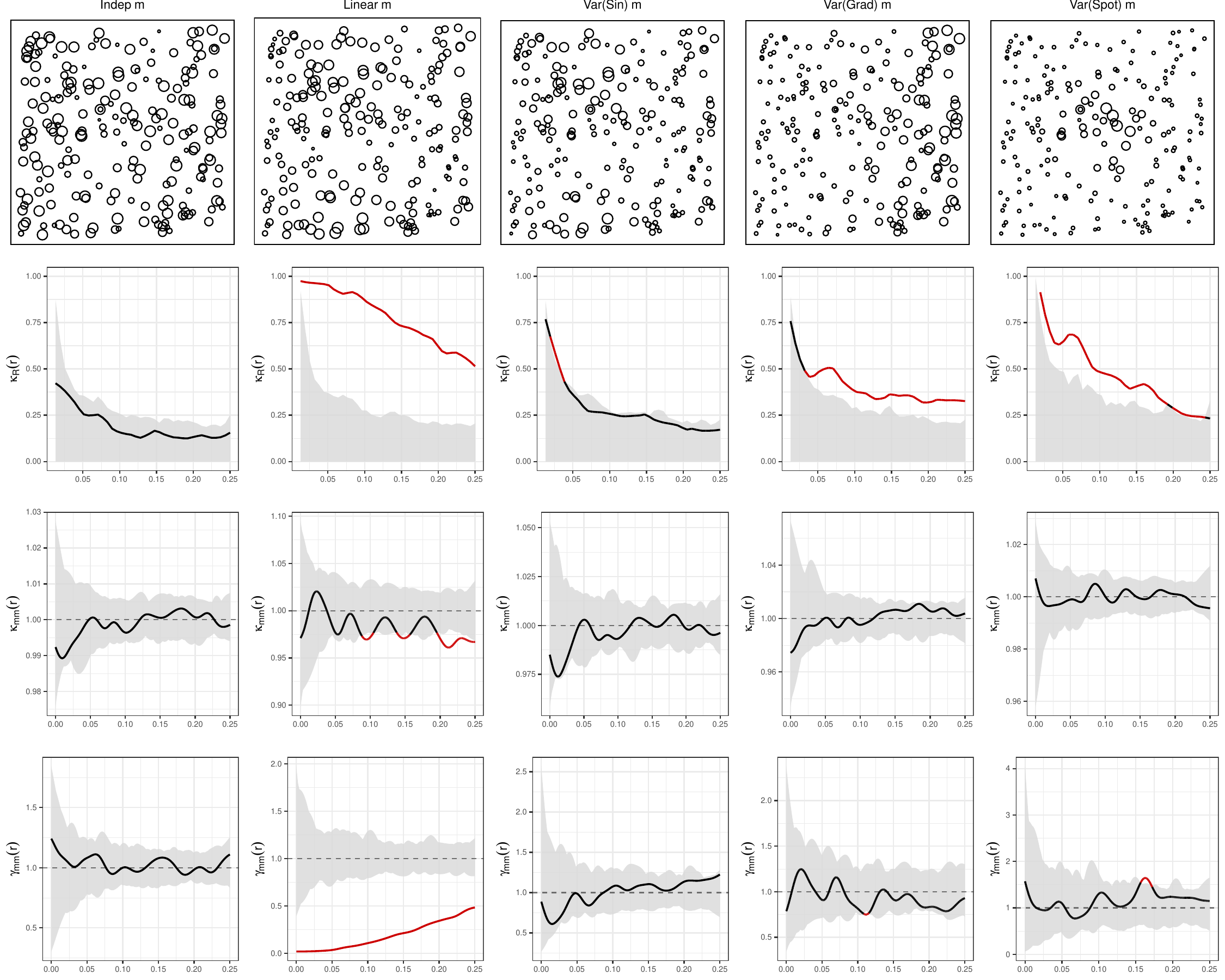} 
\end{sidewaysfigure}

A simulated point pattern together with the estimated mark distance correlation function, classical mark correlation function, and the mark variogram  
are illustrated in Figure \ref{fig:sim_univariate}. 
 
In the baseline case of random labeling 
as simulated in Scenario 1, both methods perform as expected. 
The distance correlation values remain close to zero, while the classical mark correlation mark variogram fluctuate around their respective reference values, 
confirming that none of the method identifies false patterns in the absence of mark correlation.
Other than at very small distances, the empirical mark distance correlation remains low, around $0.15$. The larger values at short distances may be due to larger uncertainty.   

When a linear field with a gradient in the mean (Scenario 2) is introduced, all 
characteristics successfully identify the spatial trend. While all 
characteristics show significant deviations from their respective independence levels, the  empirical mark correlation fluctuates around 1 except at larger distances whereas the  empirical mark distance correlation and mark variogram clearly exceed the upper or lower, respectively, envelope bound over all investigated scales. The primary advantage of the distance-based approach becomes evident in the scenarios involving spatial heteroscedasticity where the mean remains constant. In the sinusoidal variance of Scenario 3, the mark distance correlation identifies a clear and significant structural dependence. In contrast, the classical mark characteristics 
fail to react, remaining entirely within the simulation envelopes around their reference values, 
incorrectly suggesting independence between marks.  
Similarly, for Scenario 4 which is constructed using a quadratic variance gradient, the mark distance correlation is clearly above the random labeling envelope, while the mark correlation function and mark variogram remain unresponsive to this non-linear change in volatility, staying flat at the reference line. 
Finally, for the local volatility cluster (Scenario 5), the mark distance correlation detects the localized interaction, whereas the classical approaches again show no significant deviation from the null case. These results collectively demonstrate that while all methods can identify first-order trends, only the mark distance correlation is able to detect complex dependencies between the marks such as heteroscedasticity and localized variance clusters. However, since the mark distance correlation function looks very similar in all the more complex scenarios 3-5, we would not be able to distinguish between these different scenarios from $\kappa_{\mathcal{R}}(r)$ alone; Section~\ref{sec:interpretation} sets out how comparing $\kappa_{\mathcal{R}}(r)$ against the classical characteristics and inspecting conditional scatterplots can further diagnose the type of dependence detected. Therefore, when the mark distance correlation indicates correlation but the traditional mark characteristics do not, we can only say that the dependence structure is complex, e.g.\ non-linear, possibly combined with spatial inhomogeneity of marks. For example, it can be seen in Figure \ref{fig:sim_univariate} in the two rightmost point patterns on the top that the marked point patterns are inhomogeneous in terms of the marks. We stress that, as introduced in Section~\ref{sec:mmarkdcor}, $\kappa_{\mathcal{R}}(r)$ and $\gamma_{mm}(r)$ are designed to detect general stochastic dependence between marks under stationary and isotropic marking, not specifically trends, heteroscedasticity, or localized variance clusters; the present simulation study merely exploits their general sensitivity to any deviation from the random labelling hypothesis, consistent with the caveat noted at the start of Section~\ref{Simulation}.

\subsection{Multivariate marks: vectorial dependence structure}

The multivariate experiment demonstrates the capability of $k_{\mathcal{R}}(r)$ to capture dependencies between the mark components $m_{i,1}$ and $m_{i,2}$ in $\mathbf{m}(x_i) = (m_{i,1}, m_{i,2})^\top$ without dimensionality reduction. To define the underlying mark fields, we utilize the auxiliary functions $F_A(x_i) = \sin(3.5x_{i1})\cos(3.5x_{i2})$ and $G(x_i) = 2x_{i1}-1$. Following an independent reference Scenario M.1 ($m_{i,1}, m_{i,2} \sim \mathcal{U}[-1, 1]$ for all $i$) and a linearly dependent structure in Scenario M.2 ($m_{i,1} = F_A(x_{i}) + \epsilon_{i,1}$ and $m_{i,2} = F_A(x_{i}) + \epsilon_{i,2}$ with $\epsilon_{i,1},\epsilon_{i,2}\sim N(0,0.1^2)$), the study addresses complex non-linear relationships. In Scenario M.3 (Squared Dependence), a parabolic relationship is modeled by $m_{i,1} = F_A(x_{i}) + \epsilon$ and $m_{i,2} = (F_A(x_{i}))^2 + \epsilon$, resulting in an approximately zero Pearson correlation between the two mark components. Scenario M.4 (Inverse Linear) tests the detection of strong negative correlations via $m_{i,1} = F_B(x_{i}) + \epsilon$ and $m_{i,2} = -F_B(x_{i}) + \epsilon$ with $F_B(x_i)= \cos(2x_{i1}+2x_{i2})$. Finally, Scenario M.5 (V-Shape) couples the marks through $m_{i,1} = G(x_{i}) + \epsilon$ and $m_{i,2} = |G(x_{i})| + \epsilon$. This non-monotone dependency causes local correlations to cancel out globally, leading traditional mark summary characteristics to fail, whereas the mark distance characteristics approach reveals the absolute stochastic entanglement of the attributes.

To evaluate the capability of the multivariate mark distance correlation to detect complex dependence structures, we analyzed the five distinct vectorial scenarios M1 to M.5 given above using the same statistical summaries as before. 
Table \ref{tab:MultivarDcorSummary} summarizes the performance of the multivariate distance correlation function across five simulation scenarios based on 200 replications per scenario with global envelopes computed from 499 simulations. Under the independent scenario, where the null hypothesis of mark independence holds, the mean and median p-values are $0.51$ and $0.49$ respectively, with a rejection rate of $0.03$.  Only 1\% of the tests produce $p$-values below $0.01$, and the AUC of $0.49$ is consistent with the absence of any systematic dependence structure.

\begin{table}[ht]
\centering
\begin{tabular}{lccccc}
\hline
Metric & Independent & Linear & Var(Grad) & Var(Sin) & Var(Spot) \\ 
\hline
 Mean $p$ & 0.51 & 0.00 & 0.00 & 0.00 & 0.00   \\ 
  Median $p$ & 0.49 & 0.00 & 0.00 & 0.00 & 0.00   \\ 
  IQR($p$) & 0.51 & 0.00 & 0.00 & 0.00 & 0.00   \\ 
  Reject rate & 0.03 & 1.00 & 1.00 & 1.00 & 1.00  \\ 
  Prop($p<0.01$) & 0.01 & 1.00 & 1.00 & 1.00 & 1.00 \\ 
  AUC & 0.49 & 1.00 & 1.00 & 1.00 & 1.00  \\ 
\hline
\end{tabular}
\caption{Summary of multivariate mark distance correlation function  across all simulation scenarios based on 200 point patterns per scenario with corresponding global envelopes computed from 499 simulations. Heading abbreviations read as Indep = independent noise, Linear = linear field,  Var(Grad) = sinusoidal heteroscedasticity,  Var(Sin) = quadratic variance gradient,  Var(Spot) = local volatility cluster.} 
\label{tab:MultivarDcorSummary}
\end{table}

For all four dependence scenarios M.2 to M.5, the results are identical and equally striking with mean and median $p$-values 0.00, the rejection rate is 1.00, and all tests yield $p$-values below 0.01. The AUC reaches 1.00 in all cases, indicating complete separation between the observed curve and the null envelope. This demonstrates that the multivariate distance correlation function detects all forms of dependence considered here with 100\% power, including linear relationships, gradient-based effects, sinusoidal patterns, and spatially localized spot structures. Importantly, there is no moment-based method that models the joint multivariate dependence between two marks at point $i$ and two marks at point $j$ across spatial scales. Except the cross-mark correlation function, mark correlation functions commonly operate on first moments and either treat marks separately or are limited to univariate marks. The proposed approach based on distance correlation extends this to spatially indexed dependence functionals that capture the full conditional two-point mark distribution for multivariate marks. The simulation results confirm that this functional not only fills this methodological gap but also achieves excellent statistical performance with controlled maximal power across all dependence structures considered. 

\begin{sidewaysfigure}
    \centering
        \caption{Results for one simulation run: Marked point patterns (top, middle) and their mark distance correlation $k_{\mathcal{R}}(r)$ (bottom). The envelopes are based on 499 Monte Carlo permutations under random labeling. Heading abbreviations read as Indep = independent noise, Linear = linear field,  Var(Grad) = sinusoidal heteroscedasticity,  Var(Sin) = quadratic variance gradient,  Var(Spot) = local volatility cluster.}
    \label{fig:sim_multivariate}
    \includegraphics[scale=.8]{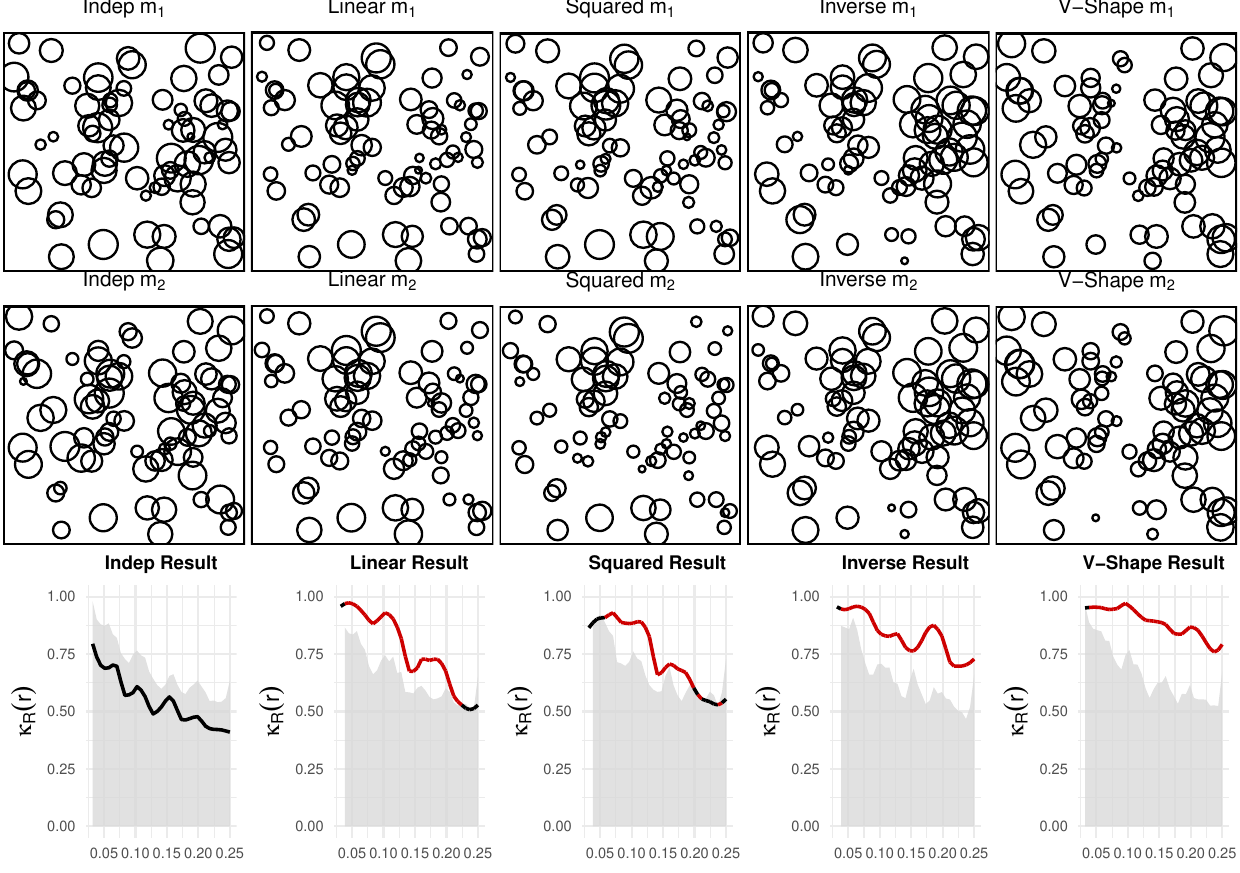}  
\end{sidewaysfigure}

The simulated patterns and resulting mark distance correlation functions for one of the patterns are presented in Figure \ref{fig:sim_multivariate} demonstrate the effectiveness of the proposed $k_{\mathcal{R}}(r)$ characteristic in capturing multi-dimensional interactions between mark components. 
In the baseline case of independent vectorial noise of Scenario M.1, the observed mark distance correlation function remain entirely within the simulation envelopes, correctly identifying a lack of spatial dependence between the two mark components. Conversely, in the presence of strong linear dependence as in Scenario M.2, the characteristic shows significantly elevated values across all spatial scales, clearly lying above the upper simulation envelope. 
The analytical power of the distance-based approach is most prominent in scenarios featuring non-linear and non-monotone relationships. In Scenario M.3, where a parabolic dependency exists between components, the mark distance correlation function  characteristic maintains high with values outside the random labeling envelope, successfully capturing the stochastic entanglement that traditional product-moment correlations would typically fail to detect. This trend continues in Scenario M.4, where the strong negative linear correlation is identified by significantly high values of the mark distance correlation function. 
Finally, in the complex case of a V-shaped dependency as simulated in Scenario M.5, the multivariate mark distance correlation function identifies the non-monotone relationship as a significant spatial structure of marks. Despite the non-linear nature of this coupling, the observed curve remains consistently above the simulation envelopes. These findings confirm that the multivariate mark distance correlation provides an efficient unified framework for detecting various forms of spatial interactions between vectorial attributes. However, as in the univariate case, even though we can detect existence of such complex dependence structures, we are not able to distinguish between them based on the mark distance correlation.


\subsection{Partial dependence structures in multivariate marks}

Since Scenarios P.2 and P.4 below are driven by a latent component $u_i$ in addition to the spatial confounder field $F_A(x_i)$, the resulting mark fields are not exactly stationary; following the caveat of Section~\ref{Simulation}, a visual check of the simulated mark maps for spatial inhomogeneity (analogous to Figure~\ref{fig:sim_univariate} for the univariate case) is therefore a useful complement to the global envelope tests reported below, and we recommend such maps be inspected alongside the test results.
To evaluate the performance of the partial mark distance correlation to detect conditional dependencies between mark components $m_{i,1}$ and $m_{i,2}$ given a third confounding or mediating variable $m_{i,3}$, we assume that each point $x_i = (x_{i1}, x_{i2})^\top$ is augmented by a trivariate mark vector $\mathbf{m}(x_i) = (m_{i,1}, m_{i,2}, m_{i,3})^\top$. The spatial dependence structures of the marks is modeled through two independent continuous random fields as auxiliary functions $F_A(x_i) = \sin(3.5x_{i1})\cos(3.5x_{i2})$ and $F_B(x_i) = \cos(2x_{i1} + 2x_{i2})$ to generate the following four simulation scenarios. As an initial dependence structure on the marks, we assume that the mark components $m_{i,1}$ and $m_{i,2}$ are conditionally independent given $m_{i,3}$ corresponding to spurious correlation or confounding (Scenario P.1).  Both $m_{i,1}$ and $m_{i,2}$ are driven solely by the shared spatial covariate field $F_A(x_i)$ with $m_{i,3} = F_A(x_i) + \epsilon_{i,3}$, $m_{i,1} = m_{i,3} + \epsilon_{i,1}$ and $m_{i,2} = m_{i,3} + \epsilon_{i,2}$ where $\epsilon_{i,1} \sim \mathcal{N}(0, 0.2^2)$ and $\epsilon_{i,2}, \epsilon_{i,3} \sim \mathcal{N}(0, 0.1^2)$ are independent Gaussian noise terms. To introduce a nonlinear conditional dependence structure (Scenario P.2), we extend the model by introducing a latent component $u_i$ in addition to the spatial confounder $m_{i,3}$. In contrast to a purely confounded setting, $u_i$ induces residual variation in $m_{i,1}$ and $m_{i,2}$ that is not fully explained by $m_{i,3}$. The data generating process is given by $m_{i,3} = F_A(x_i) + \epsilon_{i,3}$,  $m_{i,1} = 0.6\,m_{i,3} + u_i + \epsilon_{i,1}$ and $m_{i,2} = 0.3\,m_{i,3} + 2.5(m_{i,1}^2) + 1.5u_i + \epsilon_{i,2}$ where $u_i \sim \mathcal{N}(0,1)$ and $\epsilon_{i,1}, \epsilon_{i,2} \sim \mathcal{N}(0, 0.2^2)$, $\epsilon_{i,3} \sim \mathcal{N}(0, 0.3^2)$ are independent Gaussian error terms. This specification induces a nonlinear dependence between $m_{i,1}$ and $m_{i,2}$, however, the strength of the conditional effect given $m_{i,3}$ depends on the relative contribution of the latent component $u_i$ and the degree to which $m_{i,1}$ is explained by $m_{i,3}$. Consequently, a detectable but not necessarily strong residual conditional dependence may remain after conditioning on $m_{i,3}$. Next, Scenario P.3 is designed to simulate collider bias by generating  $m_{i,1}$ and $m_{i,2}$ from independent spatial fields but jointly determine $m_{i,3}$. More precisely, we induce a spurious conditional dependence in the form of $m_{i,1} = F_A(x_i) + \epsilon_{i,1}$,  $m_{i,2} = F_B(x_i) + \epsilon_{i,2}$ and $m_{i,3} = m_{i,1} + m_{i,2} + \epsilon_{i,3}$ where $\epsilon_{i,1}, \epsilon_{i,2} \sim \mathcal{N}(0, 0.2^2)$ and $\epsilon_{i,3} \sim \mathcal{N}(0, 0.1^2)$. Finally, Scenario P.4 introduces a strong residual nonlinear dependence that is not fully explained by $m_{i,3}$. The model is defined as $m_{i,3} = 0.6\,F_A(x_i) + \epsilon_{i,3}$, $m_{i,1} = 0.5\,m_{i,3} + u_i + \epsilon_{i,1}$ with $u_i \sim \mathcal{N}(0,1)$ and $m_{i,2} = 0.4\,m_{i,3} + 2(m_{i,1}^2) + 1.5u_i + \epsilon_{i,2}$, where $\epsilon_{i,1}, \epsilon_{i,2} \sim \mathcal{N}(0, 0.3^2)$ and $\epsilon_{i,3} \sim \mathcal{N}(0, 0.4^2)$ are independent Gaussian noise terms. This induces a strong nonlinear conditional dependence between $m_{i,1}$ and $m_{i,2}$ that remains detectable after conditioning on $m_{i,3}$. In contrast to Scenario P.2, the dependence structure in Scenario P.4 is dominated by the nonlinear transformation of $m_{i,1}$ and only weakly mediated by the confounding variable $m_{i,3}$. 

Table \ref{tab:pdcor_results} summarizes the performance of the partial distance correlation function across all four simulation scenarios based on 200 replications per scenario with global envelopes computed from 499 simulations.  
\begin{table}[ht]
\centering
\begin{tabular}{lcccc}
  \hline
  & Spurious & Moderate Conditional & Collider & Strong Conditional \\
  &  Confounder & Nonlinear & Bias & Signal \\
    Metric & (P.1) & (P.2) & (P.3) & (P.4) \\
  \hline
  Mean $p$          & 0.491 & 0.002 & 1.00 & 0.002 \\
  Median $p$        & 0.494 & 0.002 & 1.00 & 0.002 \\
  IQR $p$           & 0.489 & 0.000 & 0.000 & 0.000 \\
  Rejection rate    & 0.045 & 1.00 & 0.000 & 1.00 \\
  Prop($p<0.01$)    & 0.010 & 1.00 & 0.000 & 1.00 \\
  AUC               & 0.509 & 0.998 & 0.000 & 0.998 \\
  \hline
\end{tabular}
\caption{Simulation results for the partial mark distance correlation function across all simulation scenarios based on 200 point patterns per scenario with corresponding global envelopes computed from 499 simulations.}
\label{tab:pdcor_results}
\end{table}
The simulation results demonstrate that the partial mark distance correlation successfully identifies 
confounded associations and genuine conditional dependence structures. In the purely confounded setting (Scenario P.1), both mark variables are generated through the common spatial component $m_{i,3}$, such that any apparent association between $m_{i,1}$ and $m_{i,2}$ is entirely explained by the confounder. After conditioning on $m_{i,3}$, the rejection rate remains close to the nominal significance level ($4.5\%$), while the mean $p$-value is approximately 0.5. In contrast, Scenarios P.2 and P.4 introduce genuine residual conditional dependence through the nonlinear term $m_{i,1}^2$ in the data-generating process of $m_{i,2}$. Even after accounting for the confounding variable $m_{i,3}$, information about $m_{i,2}$ remains encoded in the magnitude of $m_{i,1}$, producing a nonlinear association that cannot be explained by the conditioning variable alone. This is particularly relevant because such relationships would be difficult to detect using methods that primarily target linear conditional dependence. The partial mark distance correlation, however, detects these nonlinear structures with essentially perfect power, yielding rejection rates of $100\%$ and mean $p$-values close to zero in both scenarios.

Although both scenarios P.2 and P.4 are detected almost perfectly, their underlying mechanisms differ. In Scenario P.2, the residual dependence is partially mediated through the latent variable $u_i$ and competes with the confounding effect of $m_{i,3}$. Consequently, the conditional dependence between $m_{i,1}$ and $m_{i,2}$ arises from a combination of nonlinear transformation and shared latent variation. In Scenario P.4, on the other hand, the dependence structure is dominated by the quadratic contribution of $m_{i,1}$ to $m_{i,2}$, while the influence of $m_{i,3}$ is comparatively weaker. The equally high rejection rates therefore suggest that the proposed measure remains highly sensitive across a broad range of nonlinear conditional dependence structures, from partially mediated effects to strongly persistent residual associations. The collider scenario (Scenario P.3) provides an additional and particularly challenging validation setting. Here, $m_{i,1}$ and $m_{i,2}$ are generated independently from separate spatial fields, while $m_{i,3}$ is constructed as a common consequence of both variables. Conditioning on such a collider is known to induce artificial associations between otherwise independent variables. Nevertheless, the partial mark distance correlation does not produce false detections in this setting, resulting in a rejection rate of zero and $p$-values close to  one. This behaviour indicates that the procedure is not merely responding to any dependence structure introduced through conditioning, but specifically targets residual conditional dependence that remains after accounting for the conditioning variable. Overall, the results show that the partial mark distance correlation effectively isolates genuine conditional dependence while removing associations that are fully explained by a third mark component. At the same time, it retains excellent sensitivity to nonlinear relationships, including quadratic effects and latent-variable-driven interactions, demonstrating its suitability for complex spatial mark structures that extend beyond classical linear dependence models.

\section{Applications}\label{Application}
  
\subsection{Application to forestry data}

\begin{figure}
    \centering
        \caption{
    Multivariate mark distance correlation of all three marks with global envelopes computed from $n = 499$ simulations under the random labeling hypothesis (left), point pattern with pines as discs and spruces as triangles and size corresponding to the radial increment (middle) and the height (right).}
    \label{fig:res_multi}
    \includegraphics[scale=.4]{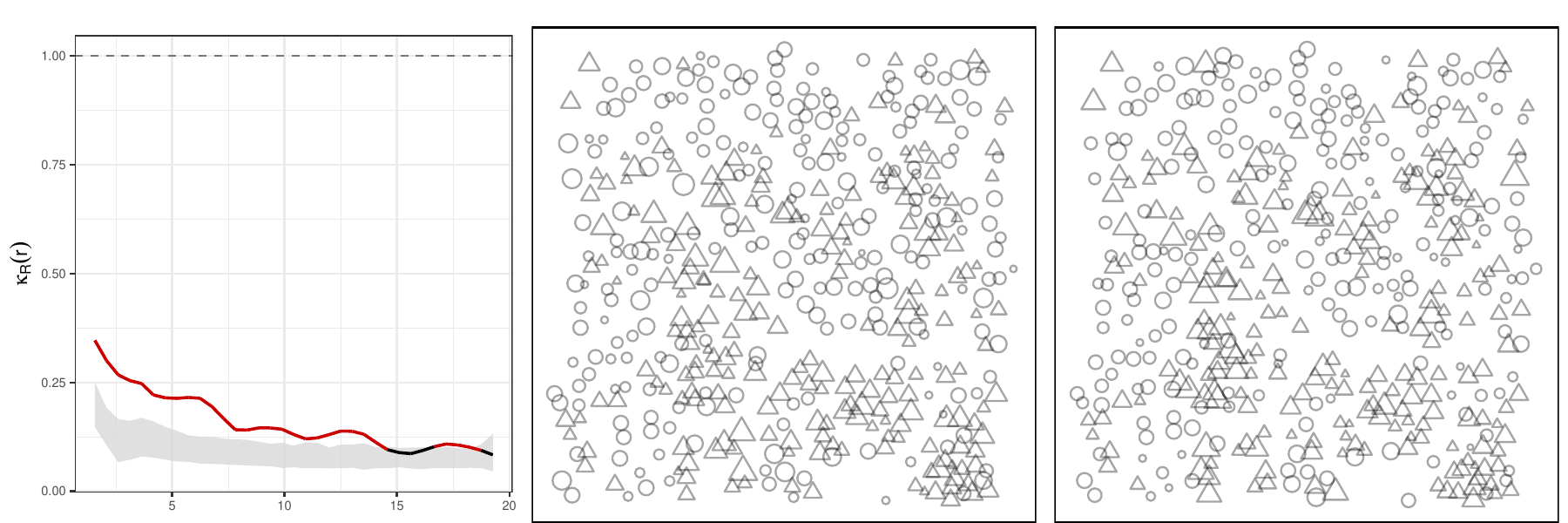} 
\end{figure}

We apply the mark distance correlation to a forestry data set from Vesijako in southern Finland \cite{Habel2019}. The pattern of trees shown in Figure \ref{fig:res_multi} (middle, right) have been observed in a mixed pine-spruce stand in a plot of size $80\text{m}\times 80 \text{m}$. In addition to the coordinates of the trees, species (236 pines and 213 spruces), 
height, and radial increment are available for each tree. 

We first analyse all three marks, namely species, height, and radial increment, using the mark distance correlation. The distance between mark vectors consisting of different types of marks is computed by using the Gower distance. The estimated multivariate mark distance  correlation function and point patterns by species are shown in Figure \ref{fig:res_multi}.
Taking all three marks jointly into account reveals a strong short-range spatial dependence in the multivariate trait space since the estimated mark distance correlation is clearly outside the envelopes constructed under random labeling. The effect is most pronounced at small spatial scales and gradually decreases with increasing distance, suggesting a pronounced local dependence structure. 

\begin{sidewaysfigure}
    \centering
        \caption{Mark correlation and mark distance correlation functions computed from the Vesijako tree data with the radial increment (i15) and height (hi) as marks by species and global envelopes computed over 499 permutations of marks. First column: mark correlation function $k_{mm}(r)$ of radial increment (top) and height (middle), and mark cross correlation function of radial increment and height (bottom) for pines. Second column is as the first column except that the function is the mark distance correlation. Third column: scatter plots for pines with $m(\circ)$ at the y-axis and $m(\mathbf{r})=r$ at the x-axis. Fourth, fifth, and sixth columns are as the first, second, and third, respectively, but for spruce.}
    \label{fig:res_markcorrdmark}
    \includegraphics[scale=.485]{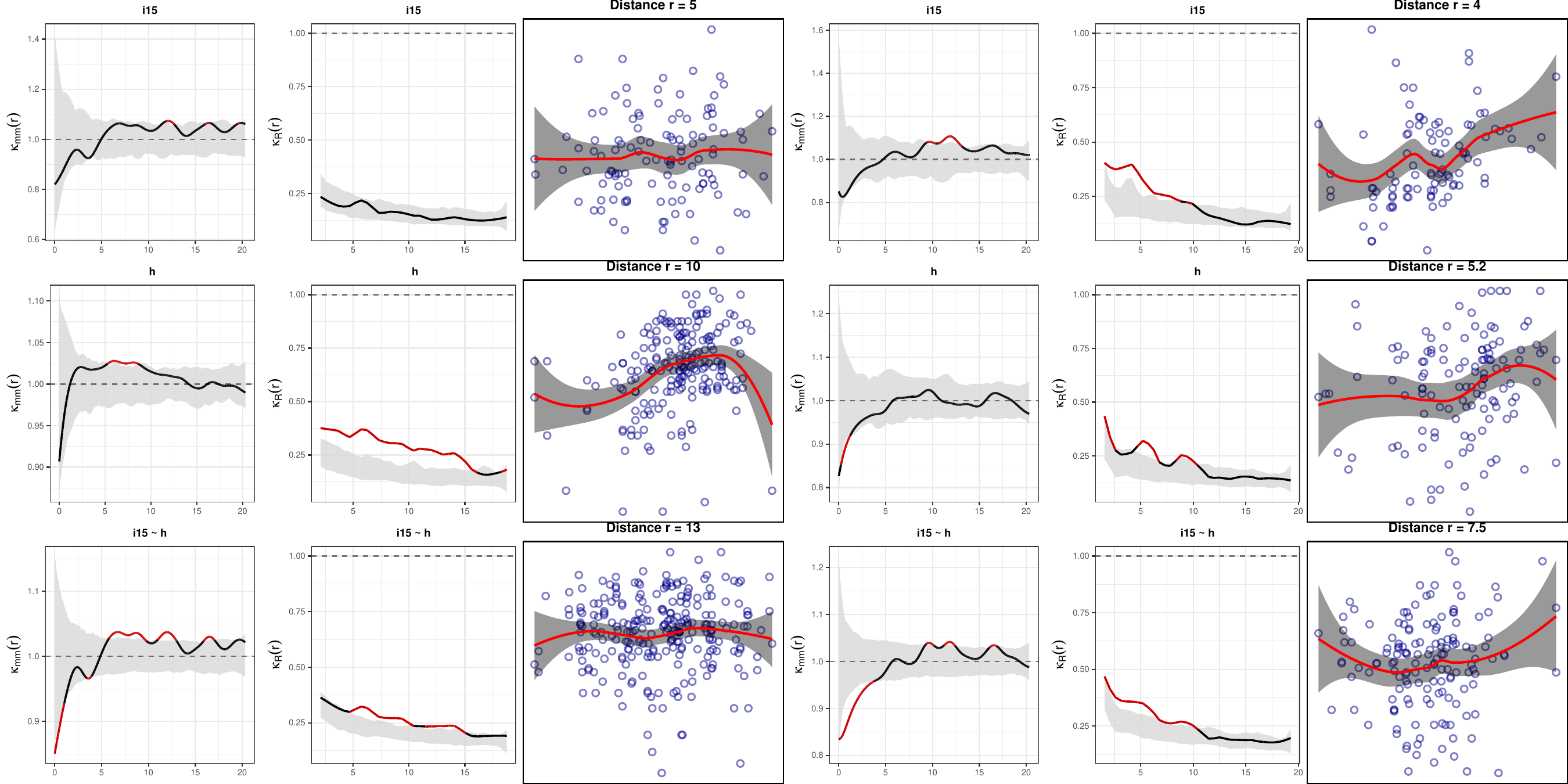} 
\end{sidewaysfigure}

To better understand the observed dependence between marks, 
we treated pines and spruces separately and computed the auto- and cross-type versions of the proposed mark distance correlation function. 
We also included Stoyan's mark correlation function for comparison. 
The results are shown in Figure \ref{fig:res_markcorrdmark}.
The "mark cross correlation" function shown in the bottom row of the first and fourth columns is the bivariate correlation of Section~\ref{Preliminaries} in the tradition of Stoyan (1987), i.e.\ $t_f(m_1(x),m_2(y)) = m_1(x)\,m_2(y)$ evaluated for radial increment against height, and is compared against its distance-based counterpart $\kappa_{\mathcal{R},ij}(r)$ in the second and fifth columns. Because the classical mark correlation/variogram panels of Figure~\ref{fig:res_markcorrdmark} are shown with pointwise (rather than global) simulation envelopes, they do not carry a single test-wide $p$-value directly comparable to the global GET-ERL test used for the mark distance correlation; the classical results are therefore described qualitatively (departures from, or containment within, the pointwise band), while $p$-values are reported for the distance-based panels only.

For pines (the first three columns), the classical mark correlation function reveals only weak departures from the pointwise null band for radial increment (i15) around 12m and height (h) around 5-7m, and somewhat stronger correlation at several distances between them. The mark distance correlation function, by contrast, does not detect correlation for the radial increment ($\kappa_{\mathcal{R}}(r)$, $p=0.40$) but indicates significant correlation up to 15m or so for height ($p=0.002$) and for the radial-increment-height cross-correlation ($p=0.002$), exceeding what is indicated by the classical mark correlation function. The scatter plot at distance 10m (third column, second row) indicates non-linear dependence between the heights which may explain why the mark correlation function does not recognize it.

For spruces (the last three columns), the mark correlation function shows departures from the pointwise null band for radial increment and their cross-correlation around 12m while the empirical curve for height remains within the pointwise band except at very short distances. The mark distance correlation function, on the other hand, is significant for radial increment, height, and their cross-correlation alike (all $p=0.002$), revealing mark correlation for height across several distances that the classical statistic misses.
The scatter plots reveal some non-linear dependence between radial increments at distance 4m, between heights at distance 5.2m and between the two marks at distance 7.5m, all of which were captured by the distance correlation function but not by the mark correlation function. 

\begin{figure}
    \centering
        \caption{Mark distance correlation computed from the Vesijako tree data using the Gower distances with the radial increment (i15), height (hi) and species as marks and global envelopes computed over 499 simulations. Mark distance correlation functions (first row), distance cross-correlation functions (second row), partial distance auto-correlation functions  (third row), and partial distance cross-correlation functions (fourth row).}
    \label{fig:res_dmarksGower}
    \includegraphics[scale=.38]{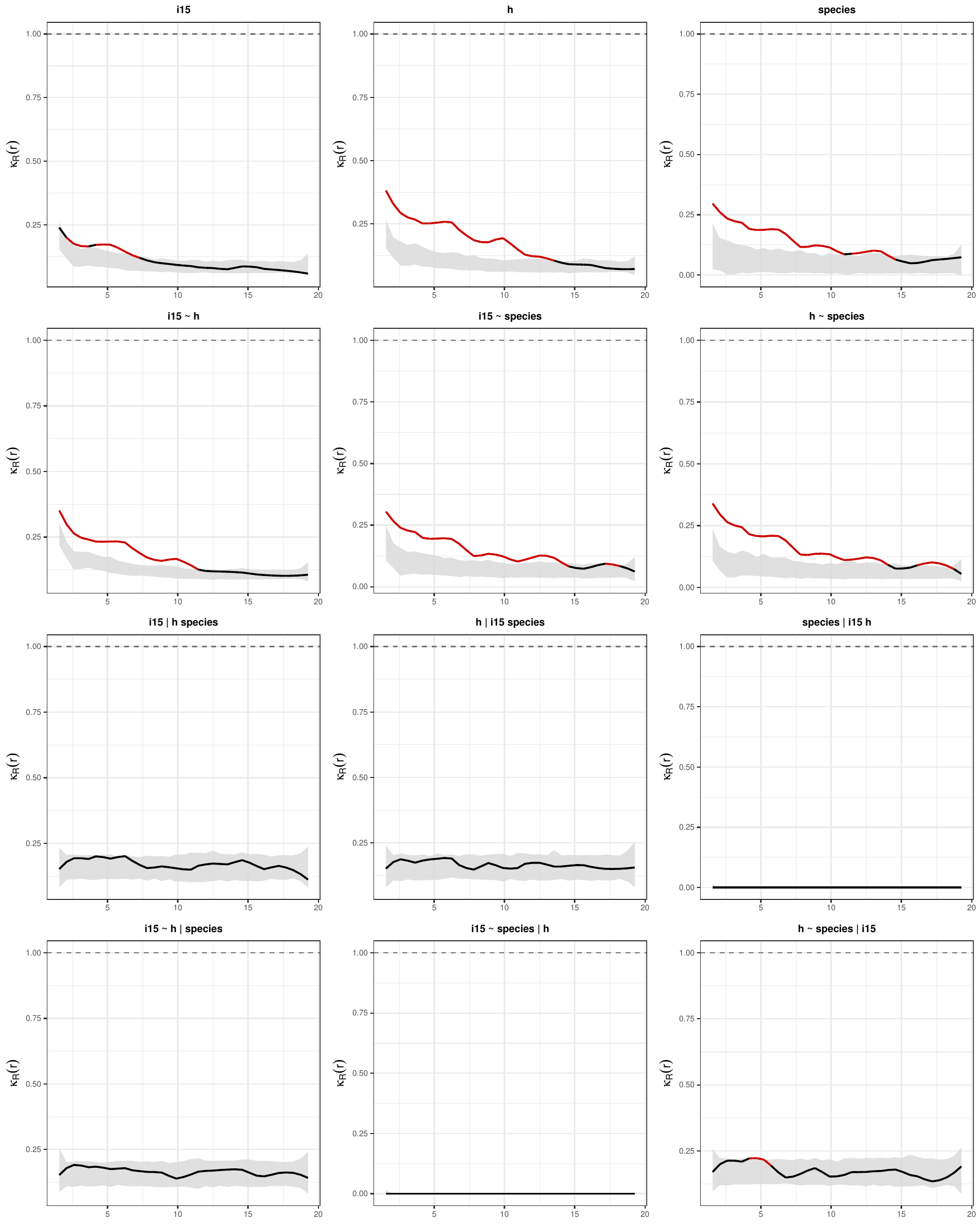}
\end{figure}

Finally, we computed all the mark distance correlation, distance cross-correlation, partial distance auto-correlation, and partial distance cross-correlation functions including all three marks using the Gower distance between the mark vectors shown in Figure \ref{fig:res_dmarksGower}. 
We can see significant distance correlation between radial increments around 5m, between heights up to 12m, and between species up to 10m. The cross-correlations between radial increment and height, radial increment and species, and between height and species are also significant up to 10-12m. Among the partial correlations, only the cross-correlation between height and species, when the effect of radial increment has been removed, is significant, with the most prominent deviation around 5m (Figure \ref{fig:res_dmarksGower}, bottom right).  
Since the unconditional cross correlation between height and species was also significant (Figure \ref{fig:res_dmarksGower}, right column on the second row from the top), this indicates that the association between the two marks is not merely an artifact of both being driven by the same growth related component captured through radial increment. Instead, height and species remain linked at that spatial scale even once the shared growth variation has been accounted for, pointing to a genuine, direct short range association between tree height and species identity rather than one mediated through stem increment.

\subsection{Application to Spanish population data}
 
\begin{figure}
    \centering
        \caption{Point pattern with local population trajectory as mark for the Toledo data. See text for details.}
    \label{fig:toledo}
    \includegraphics[width=0.65\linewidth]{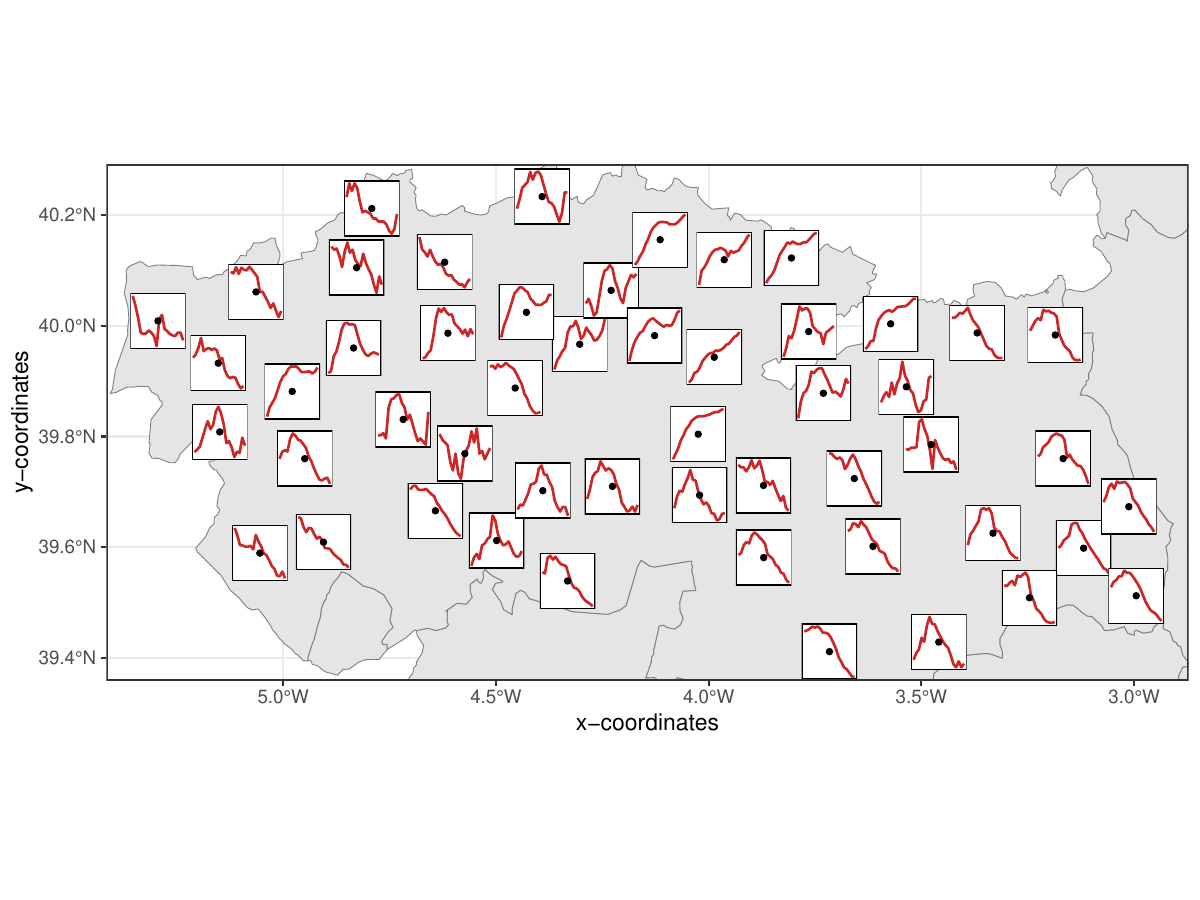}

\end{figure}

To illustrate the functional-mark extension of the mark distance characteristics introduced above, we analyse yearly population totals for the municipalities of the province of Toledo, Spain, obtained from the Spanish National Statistics Institute (INE). Each municipality is represented by a point located at its centroid, projected to the UTM zone 30N coordinate reference system (EPSG:25830) so that pairwise distances are expressed in metres. The mark attached to point $x_i$ is the municipality's population trajectory $g(x_i)(t)$, $t = 2003, \ldots, 2022$, a curve of $p = 20$ annual population totals evaluated on a common yearly grid $\mathcal{T} = \{2003, \ldots, 2022\}$. One municipality (Retamoso de la Jara) was excluded for lack of geocoded coordinates, leaving $n = 203$ points. A visualisation of the point pattern with the corresponding function-valued marks is shown in Figure \ref{fig:toledo}.

We compute the mark distance covariance $\kappa_{\mathcal{V}}(r)$ and correlation $\kappa_{\mathcal{R}}(r)$ functions using the $L^2$ distance
\[
d(g_i, g_j) = \left( \int_{\mathcal{T}} |g_i(t) - g_j(t)|^2 \, \mathrm{d}t \right)^{1/2},
\]
approximated by the trapezoidal rule on the yearly grid, in place of the Euclidean distance used for scalar or vector marks. Statistical significance is assessed against the random labelling hypothesis using a one-sided global envelope test (extreme rank length) based on 499 permutations of the population trajectories over the fixed municipality locations.

\begin{figure}
    \centering
        \caption{Mark distance covariance $\kappa_{\mathcal{V}}(r)$ (left) and mark distance correlation $\kappa_{\mathcal{R}}(r)$ (right) for municipal population trajectories (2003--2022) in the province of Toledo, Spain, computed via the $L^2$ distance between curves. Grey ribbons show one-sided global (ERL) envelopes based on 499 random-labelling permutations; red segments indicate distances at which the observed curve exceeds the envelope.}
    \label{fig:toledo_func}
    \includegraphics[scale=.65]{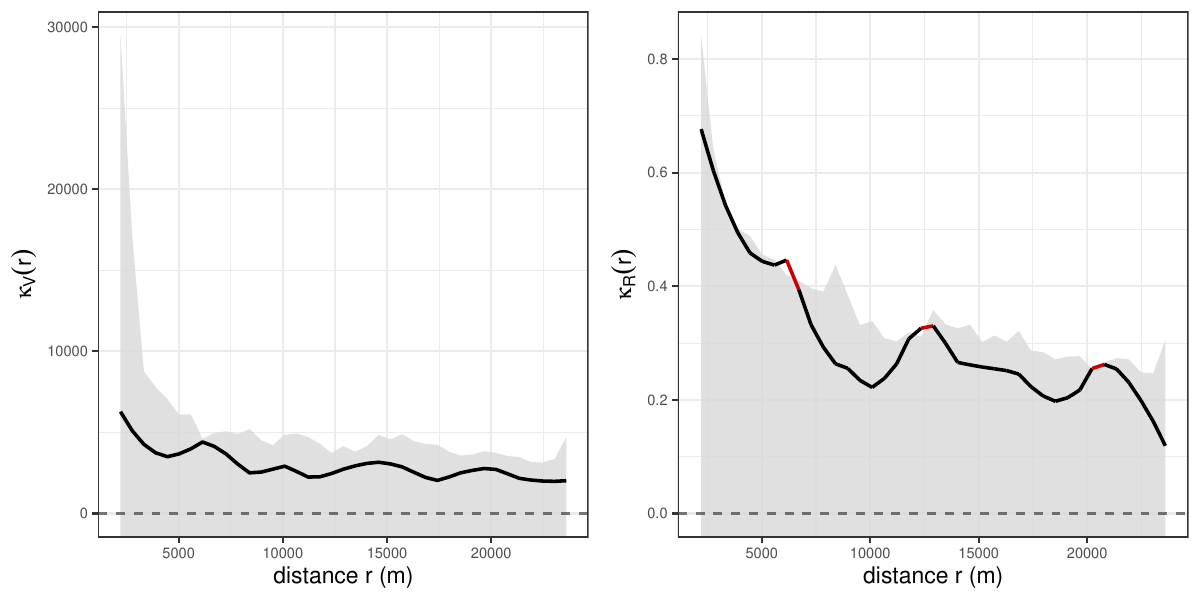}
\end{figure}

Results are shown in Figure~\ref{fig:toledo_func}. The mark distance correlation $\kappa_{\mathcal{R}}(r)$ exceeds the upper simulation envelope up to a distance of approximately 7\,km ($p = 0.026$), indicating that neighbouring municipalities exhibit significantly associated population trajectories at short spatial range. In contrast, the unnormalised mark distance covariance $\kappa_{\mathcal{V}}(r)$ remains within its envelope throughout ($p = 0.182$). This discrepancy is expected: the covariance is driven by the raw magnitude of population change, which is dominated by the small number of larger municipalities (including the provincial capital), whereas the normalised correlation is sensitive to the overall shape of the trajectories -- e.g.\ shared demographic decline or growth patterns among rural municipalities of comparable size -- irrespective of their absolute population level. This example illustrates that the functional mark distance correlation can reveal spatially structured similarity in curve-valued attributes that a scale-sensitive covariance measure may fail to detect.

 \section{Conclusion}
\label{Conclusion}
 
We have adapted the distance correlation of \cite{10.1214/009053607000000505} to marked point processes, yielding mark summary characteristics that address two distinct limitations of the classical mark correlation toolkit. First, for a single quantitative mark, the proposed characteristic $\kappa_{\mathcal{R}}(r)$ recognises non-linear and non-monotone association among marks that remain invisible to moment-based test functions. Second, for a multivariate mark vector, it assesses joint dependence between all mark components directly, without requiring dimension reduction or a pairwise decomposition into scalar comparisons. 

Our simulation study demonstrates that, already for a single quantitative mark, the mark distance correlation $\kappa_{\mathcal{R}}(r)$ is uniquely capable of detecting complex stochastic dependencies that remain invisible to traditional metrics. A decisive advantage of this framework is its departure from the strict requirement of positive real-valued marks: while classical product-based functions like Stoyan's $\kappa_{mm}(r)$ lose interpretability or require artificial transformations when applied to centred data or residuals, our distance-based approach is invariant to translation and remains numerically robust for marks covering negative value ranges. This global sensitivity to a general departure from independence represents a significant improvement over the linear or monotonic focus of the classical mark characteristics.

For multivariate marks, the ability of assessing dependence between all mark components jointly,  was illustrated for species, height and radial increment in the forestry application of Section~\ref{Application}. For a single function-valued mark, our framework further avoids the loss of information inherent in the pointwise integration of classical test functions: by utilising the global $\mathcal{L}_2$-geometry of the mark space, $\kappa_{\mathcal{R}}(r)$ captures shape variation and phase-locked dependence that would otherwise be masked by the sign or the linearity of the underlying process, as illustrated for the population trajectories of Section~\ref{Application}. Extending this construction to several function-valued marks jointly, combined for instance via a suitably weighted fusion of their respective $\mathcal{L}_2$ metrics, is conceptually straightforward within the proposed framework but is not empirically demonstrated here; we leave a worked example of such genuinely multivariate functional marks, together with a systematic study of the resulting characteristic's power, for future work.

A newly introduced, more complex class of marked point processes includes cases where marks are constrained arrays or structured quantities, such as composition-valued or graph-valued marks \cite{EckardtMariCMSPP, Eckardt2023MultiFunctionMarks}. The current mark distance correlation approach can be adjusted to such marks, but this requires careful consideration of distance metrics specific to these marks.


Ultimately, this work provides two complementary tools, sensitive to non-linear dependence for a single mark and to joint dependence across multivariate mark vectors, within a single, coherent estimation framework validated by rigorous Monte Carlo inference. Together, they make these characteristics particularly suitable for complex datasets in ecology, forestry, and environmental sciences, where non-linear dependencies, multivariate mark vectors, and non-Gaussian noise structures are prevalent.

\section*{Acknowledgments}
AS gratefully acknowledges financial support from 
the Swedish Research Council (2025-04712).
We thank Risto Ojansuu for the Vesijako data.

\bibliographystyle{ecta} 
\bibliography{DmarksArxiv}

\end{document}